\documentclass[12pt,a4paper,final]{article}
\pdfoutput=1
\usepackage[title,titletoc,toc]{appendix}
\usepackage{color}
\usepackage{hyperref} 
\usepackage[utf8]{inputenc}
\usepackage{amsmath}
\usepackage{amsfonts}
\usepackage{amssymb}
\usepackage{enumitem}

\usepackage{float}

\usepackage{textcomp}
\usepackage{graphicx}
\graphicspath{ {images/} }

\usepackage{pgfplots}
\pgfplotsset{compat=1.15}
\usepackage{mathrsfs}
\usetikzlibrary{arrows}
\usepackage[font=small,labelfont=bf]{caption}

\hypersetup{ colorlinks=true, linkcolor=blue, citecolor=red }

\begin{document}

%\definecolor{qqqqff}{rgb}{0,0,1}%For tikz

\title{
\vspace{1.5cm}
Quivers for coloured planar $\phi^n$ theories at all loop orders for scattering processes from Feynman diagrams.}
\author{{Prafulla Oak\footnote {prafullaoakwork@gmail.com} }  \\ \\
  \small Indian Association for the Cultivation of Sciences\\  
  \small 2A and 2B Raja S.C. Mullick road, Kolkata-700032 \\
  \vspace{.5cm}  
  \small India \\
  \vspace{.5cm}
}
\date{}
\maketitle
\begin{abstract}
We introduce Feynman-like rules to compute quivers for two loops and higher for the coloured planar $\phi^3$ theory for winding number zero. We demonstrate this for a few cases.
Then we extend this further to the case of $\phi^n$ theories, for any n. We also construct the quivers for mixed theories. We do this at all loop order.
\end{abstract}

\tableofcontents 
\newpage

\section{Introduction}

Scattering amplitudes are the most basic observables in physics. Also, they are the only known observable of quantum gravity in asymtotically flat space-time. The scattering amplitudes in the case of AdS spaces are easily calculable as all the in and out states are observables of the local theory of the boundary, which for the AdS case is the local quantum field theory in flat space-time. This question is very hard to answer in bulk flat space-time, the reason being that we do not have a good answer to what the theory at infinity should be in this case. The initial S-matrix theorists were looking for a first principle derivation of the fundamental analyticity properties encoding unitarity and causality in the S-matrix, and in this way to find the principles for the theory of S-matrix. This approach did not succeed and a new startegy was needed to address this problem. The modern strategy is to look for new principles and laws, which one expects, to be associated with entirely new mathematical structures, that produce the S-matrix as the answer to an entirely different class of questions and see unitarity and causality as emerging from them rather than being imposed apriori. This approach was initially taken by \cite{Arkani-Hamed:2013jha} to analyse scattereing amplitudes in super-symmetric quantum field theories. They later extended it to a class of non-super-symmetric coloured scalar $\phi^3$ theories. They started by introducing a planar scattering form, which is a differential form on the space of kinematic variables, that encodes information about on-shell tree level scattering, thus "upgrading" amplitudes to forms. Further, a precise connection was established between this scattering form and a polytope known as the "Associahedron". This connection leads one precisely to a new understanding of locality and unitarity as emergent properties from the combinatorial and geometric properties of the polytopes, as one had hoped for, in the begining. To obtain these forms one looks at the duals to the Feynman diagram, as will be shown later, and then triangulates them. One can associate quivers to these triangulations. We will show this in one of the sections. The quivers for tree level processes were shown to be $A_n$ Dynkin type quivers. Similarly, the quivers for 1-loop processes were shown to be of the $D_n$ type Dynkin quivers \cite{Arkani-Hamed:2019vag}. Some of these were demonstrated by \cite{rdst} and there were unique such quivers for tree level and 1-loop level processes. For higher loops the association of Dynkin-type quivers with loop order breaks down. For 2-loops and higher there were problems with non-uniqueness of the quivers as described by \cite{rdst}, that, one could associate many quivers to a process, as there were many triangulations one could associate to each Feynman diagram. This happens because the winding number of the diagonals is undetermined and that leads to there being a non-unique set of triangulations.

A lot of progress has been made in this area parallely. Much work has been done on effective field theories \cite{Arkani-Hamed:2020blm,Cheung:2016drk}. It has also been shown that the Yangian symmetry is manifest in this approach \cite{Arkani-Hamed:2013jha,Arkani-Hamed:2014dca,Arkani-Hamed:2013kca}. It was shown in \cite{Arkani-Hamed:2017mur} that the CHY formula gives scattering amplitude for a large class of theories as integrals of world sheet moduli space. It has been known for sometime that the compactification of this moduli space is an associahedron \cite{Devadoss,deligne} and in \cite{Arkani-Hamed:2017mur} it was shown that this worldsheet associahedron is diffeomorphic to the associahedron sitting inside the kinematic space. Scattering equations, which are the basic building blocks of the CHY formula, are precisely these diffeomorphisms. It naturally follows, that the CHY integrand for the $\phi^3$ theory is a pullback of the canonical form on the associahedron.

Scattering amplitudes are observables of singular importance for many reasons. Computing them is one of the most important problems in physics. They are traditionally computed with the help of Feynman diagrams  by doing an order by order summation of perturbations in the coupling strength. An alternate approach was developed by Nima Arkani-Hamed et al\cite{Arkani-Hamed:2013jha}. For color planar scalar theories \cite{Arkani-Hamed:2017mur} they start by drawing a dual polygon to the Feyman diagram.

\begin{figure}[H]\label{schannel}
\centering
\definecolor{uuuuuu}{rgb}{0.26666666666666666,0.26666666666666666,0.26666666666666666}
\definecolor{zzttqq}{rgb}{0.6,0.2,0}
\begin{tikzpicture}[line cap=round,line join=round,>=triangle 45,x=1cm,y=1cm]
%\clip(-9.792871569624667,-3.0784380668456452) rectangle (6.9865788160233455,7.356782030133528);
\fill[line width=2pt,color=zzttqq,fill=zzttqq,fill opacity=0.10000000149011612] (-3.3048174205074345,1.7156906147680617) -- (-1.3232442321070987,1.7316710437067742) -- (-1.3392246610458112,3.71324423210711) -- (-3.320797849446147,3.697263803168398) -- cycle;
\draw [line width=2pt,color=zzttqq] (-3.3048174205074345,1.7156906147680617)-- (-1.3232442321070987,1.7316710437067742);
\draw [line width=2pt,color=zzttqq] (-1.3232442321070987,1.7316710437067742)-- (-1.3392246610458112,3.71324423210711);
\draw [line width=2pt,color=zzttqq] (-1.3392246610458112,3.71324423210711)-- (-3.320797849446147,3.697263803168398);
\draw [line width=2pt,color=zzttqq] (-3.320797849446147,3.697263803168398)-- (-3.3048174205074345,1.7156906147680617);
\draw [line width=2pt] (-3.320797849446147,3.697263803168398)-- (-1.3232442321070987,1.7316710437067742);
\draw [line width=2pt] (-3.9600150069946443,2.738438066845654)-- (-2.857365410223489,2.562653348519818);
\draw [line width=2pt] (-2.857365410223489,2.562653348519818)-- (-2.6975611208363652,1.204316888729267);
\draw [line width=2pt] (-2.857365410223489,2.562653348519818)-- (-2.0263831054104444,3.1539292192521753);
\draw [line width=2pt] (-2.0263831054104444,3.1539292192521753)-- (-2.1062852501040066,4.640109110552425);
\draw [line width=2pt] (-2.0263831054104444,3.1539292192521753)-- (-0.41235978260049283,2.9142227851714897);
\draw (-9.5019043957427593,4.240598387084617) node[anchor=north west] {\parbox{3.5568686301939825 cm}{Diagonal dual to the  internal propagator.}};
\draw [->,line width=2pt] (-5.590018758743309,3.8091268057393832) -- (-2.934199711629862,3.3168512355571735);
\draw [->,line width=2pt] (-5.590018758743309,3.8091268057393832) -- (-2.934199711629862,3.3168512355571735);
\begin{scriptsize}
\draw [fill=uuuuuu] (-1.3392246610458112,3.71324423210711) circle (2.5pt);
\draw [fill=uuuuuu] (-3.320797849446147,3.697263803168398) circle (2.5pt);
\end{scriptsize}
\end{tikzpicture}
\caption{The s-channel 2-2 scattering process and its dual }
\end{figure}

Here the sides of the polygon are dual to the external lines of the Feynman diagram and each diagonal is the dual to an internal propagator. Then you compute the top form for this geometry and the coefficient of that is the scattering amplitude. For example, in the case shown above, this is a 2-2 scattering in the s-channel. There is another diagram for the t-channel.

\begin{figure}[H]\label{tchannel}
\centering
\definecolor{uuuuuu}{rgb}{0.26666666666666666,0.26666666666666666,0.26666666666666666}
\definecolor{zzttqq}{rgb}{0.6,0.2,0}
\begin{tikzpicture}[line cap=round,line join=round,>=triangle 45,x=1cm,y=1cm]
%\clip(-7.140120365798409,-5.379619834020228) rectangle (9.703251735604447,5.0556002629589445);
\fill[line width=2pt,color=zzttqq,fill=zzttqq,fill opacity=0.10000000149011612] (-2.090304821165294,-0.058136997429011475) -- (-0.2525554932133697,-0.01019571061287439) -- (-0.30049678002950697,1.8275536173390496) -- (-2.138246107981431,1.7796123305229128) -- cycle;
\draw [line width=2pt,color=zzttqq] (-2.090304821165294,-0.058136997429011475)-- (-0.2525554932133697,-0.01019571061287439);
\draw [line width=2pt,color=zzttqq] (-0.2525554932133697,-0.01019571061287439)-- (-0.30049678002950697,1.8275536173390496);
\draw [line width=2pt,color=zzttqq] (-0.30049678002950697,1.8275536173390496)-- (-2.138246107981431,1.7796123305229128);
\draw [line width=2pt,color=zzttqq] (-2.138246107981431,1.7796123305229128)-- (-2.090304821165294,-0.058136997429011475);
\draw [line width=2pt] (-2.090304821165294,-0.058136997429011475)-- (-0.30049678002950697,1.8275536173390496);
\draw [line width=2pt] (-3.0491305574880374,1.3001994623615394)-- (-1.626872381942635,1.1883364597905528);
\draw [line width=2pt] (-1.626872381942635,1.1883364597905528)-- (-1.5949115240652103,2.5946142063972406);
\draw [line width=2pt] (-1.626872381942635,1.1883364597905528)-- (-0.9397139375780024,0.6609823048130449);
\draw [line width=2pt] (-0.9397139375780024,0.6609823048130449)-- (-0.9876552243941396,-0.45764772089682054);
\draw [line width=2pt] (-0.9397139375780024,0.6609823048130449)-- (0.14695523025444,0.9965713125260045);
\draw [line width=2pt] (-2.090304821165294,-0.058136997429011475)-- (-0.2525554932133697,-0.01019571061287439);
\begin{scriptsize}
\draw [fill=uuuuuu] (-0.30049678002950697,1.8275536173390496) circle (2.5pt);
\draw [fill=uuuuuu] (-2.138246107981431,1.7796123305229128) circle (2.5pt);
\end{scriptsize}
\end{tikzpicture}
\caption{The t-channel 2-2 scattering process and its dual}
\end{figure}

There is an alternate way for computing the scattering amplitude of an n-point function, for any n, at one go without looking at all the channels one by one. This is only known at the tree level. This computation is done by starting with the quiver for any n-point tree level process and drawing its Auslander-Reiten walk \cite{Barmeier:2021iyq}.

\begin{figure}[H]\label{quiverdemo}
\centering
\definecolor{xdxdff}{rgb}{0.49019607843137253,0.49019607843137253,1}
\definecolor{uuuuuu}{rgb}{0.26666666666666666,0.26666666666666666,0.26666666666666666}
\definecolor{zzttqq}{rgb}{0.6,0.2,0}
\begin{tikzpicture}[line cap=round,line join=round,>=triangle 45,x=1cm,y=1cm]
%\clip(-8.354632965140553,-5.171874257816963) rectangle (7.274226536920162,5.26334583916221);
\fill[line width=2pt,color=zzttqq,fill=zzttqq,fill opacity=0.10000000149011612] (-6.181294629475665,-0.0900978553064353) -- (-4.263643156830178,-0.07411742636772294) -- (-3.6862545535699325,1.7546157267942275) -- (-5.247060244683758,2.868854542863368) -- (-6.789079814886619,1.728758849616589) -- cycle;
\fill[line width=2pt,color=zzttqq,fill=zzttqq,fill opacity=0.10000000149011612] (-1.5150093793716473,0.14960857877425016) -- (0.32273994858027716,0.2135302945290996) -- (0.8298425580171442,1.9810866646726693) -- (-0.694500121519033,3.0095748626979364) -- (-2.1436983174113466,1.877659155962114) -- cycle;
\draw [line width=2pt,color=zzttqq] (-6.181294629475665,-0.0900978553064353)-- (-4.263643156830178,-0.07411742636772294);
\draw [line width=2pt,color=zzttqq] (-4.263643156830178,-0.07411742636772294)-- (-3.6862545535699325,1.7546157267942275);
\draw [line width=2pt,color=zzttqq] (-3.6862545535699325,1.7546157267942275)-- (-5.247060244683758,2.868854542863368);
\draw [line width=2pt,color=zzttqq] (-5.247060244683758,2.868854542863368)-- (-6.789079814886619,1.728758849616589);
\draw [line width=2pt,color=zzttqq] (-6.789079814886619,1.728758849616589)-- (-6.181294629475665,-0.0900978553064353);
\draw [line width=2pt,color=zzttqq] (-1.5150093793716473,0.14960857877425016)-- (0.32273994858027716,0.2135302945290996);
\draw [line width=2pt,color=zzttqq] (0.32273994858027716,0.2135302945290996)-- (0.8298425580171442,1.9810866646726693);
\draw [line width=2pt,color=zzttqq] (0.8298425580171442,1.9810866646726693)-- (-0.694500121519033,3.0095748626979364);
\draw [line width=2pt,color=zzttqq] (-0.694500121519033,3.0095748626979364)-- (-2.1436983174113466,1.877659155962114);
\draw [line width=2pt,color=zzttqq] (-2.1436983174113466,1.877659155962114)-- (-1.5150093793716473,0.14960857877425016);
\draw [line width=2pt] (-6.789079814886619,1.728758849616589)-- (-4.263643156830178,-0.07411742636772294);
\draw [line width=2pt] (-5.247060244683758,2.868854542863368)-- (-4.263643156830178,-0.07411742636772294);
\draw [line width=2pt] (-6.964335647472572,0.7728453073840323)-- (-5.989529482211116,0.5650997311807716);
\draw [line width=2pt] (-5.989529482211116,0.5650997311807716)-- (-5.701881761314293,-0.825197586487204);
\draw [line width=2pt] (-5.989529482211116,0.5650997311807716)-- (-5.382273182540045,1.4919646096260886);
\draw [line width=2pt] (-5.382273182540045,1.4919646096260886)-- (-6.596785781882186,2.8982423562327764);
\draw [line width=2pt] (-5.382273182540045,1.4919646096260886)-- (-4.4554083040947265,1.6038276121970751);
\draw [line width=2pt] (-4.4554083040947265,1.6038276121970751)-- (-4.247662727891465,2.866281498355352);
\draw [line width=2pt] (-4.4554083040947265,1.6038276121970751)-- (-3.560504283526833,0.629021446935621);
\draw [line width=2pt] (-2.1436983174113466,1.877659155962114)-- (0.32273994858027716,0.2135302945290996);
\draw [line width=2pt] (-0.694500121519033,3.0095748626979364)-- (0.32273994858027716,0.2135302945290996);
\draw [line width=2pt] (-2.0743243922265844,1.0445125993421451)-- (-1.7706962423910493,1.124414744035707);
\draw [line width=2pt] (-1.5469702372490757,2.6105946353359566)-- (-1.3552050899845272,2.3868686301939834);
\draw [line width=2pt] (0.019111798744738156,2.4188294880714083)-- (0.16293565919314965,2.6904967800295188);
\draw [line width=2pt] (0.43460295115126024,1.2362777466066934)-- (0.7062702431093708,1.1883364597905564);
\draw [line width=2pt] (-0.7479487903134564,0.3733345839162259)-- (-0.7479487903134564,0.10166729195811579);
\draw [line width=2pt,dash pattern=on 1pt off 1pt] (-1.1029913125656623,1.175484450461055)-- (-0.21172906017512827,1.6826025545701735);
\begin{scriptsize}
\draw [fill=uuuuuu] (-3.6862545535699325,1.7546157267942275) circle (2.5pt);
\draw [fill=uuuuuu] (-5.247060244683758,2.868854542863368) circle (2.5pt);
\draw [fill=uuuuuu] (-6.789079814886619,1.728758849616589) circle (2.5pt);
\draw [fill=uuuuuu] (0.8298425580171442,1.9810866646726693) circle (2.5pt);
\draw [fill=uuuuuu] (-0.694500121519033,3.0095748626979364) circle (2.5pt);
\draw [fill=uuuuuu] (-2.1436983174113466,1.877659155962114) circle (2.5pt);
\draw [fill=xdxdff] (-1.1029913125656623,1.175484450461055) circle (2.5pt);
\draw [fill=xdxdff] (-0.21172906017512827,1.6826025545701735) circle (2.5pt);
\end{scriptsize}
\end{tikzpicture}
\caption{Quiver for a tree level 5-point process for a $\phi^3$ interaction.}
\end{figure}

On the left is the five point tree level Feynman diagram with its dual. On the right only the dual pentagon is shown. There are two types of nodes shown in the figure, dashes and dots. The dashes on the sides of the polygon are called frozen nodes and the dots are called unfrozen nodes. The quiver corresponding to this process is given by joining the two unfrozen nodes. One can assign a direction to the line joining the two nodes by selecting a clockwise or an anticlockwise direction. We have not shown that here.

To obtain the scattering amplitude we look at the Auslander-Reiten walk of this quiver and using the techniques in \cite{Arkani-Hamed:2019vag} compute the full scattering amplitude of any n-point process. It is known how to compute tree level and 1-loop quivers for $\phi^3$ theory from \cite{rdst}. 

In this paper we propose a way to construct the quivers corresponding to each Feynman diagram at winding number zero. We do not know the answers for higher winding numbers or any consequences thereof, but nonetheless this seems like an interesting heuristic exercise to carry out. In section 2, we will give a very quick review of the "Amplituhedron" program for the coloured scalar $\phi^3$ theory. Then in section 3 we will briefly mention the known quivers for tree level and 1-loop processes. In the next section we motivate and prescribe a set of Feynman-like rules to construct quivers from Feynman diagrams directly. In section 5 we demonstrate the construction of quivers for 2-loop and 3-loop processes. After that we juxtapose the two approaches of computing the quivers. We also demonstrate the construction of quivers using our approach, step by step. Then we will see the construction of quivers for a few non-trivial cases of scattering processes. In section 6 we will prescribe Feynman-like rules to construct quivers for $\phi^{n}$ theories, $n\geq4$. In the end we summarize and then indicate possible future directions.

\section{A quick review of the "Amplituhedron" formalism for scalar theories}

In this section we summarize the results of \cite{Arkani-Hamed:2017mur}.

\subsection{Kinematic space}

The Kinematic space $\mathcal{K}_n$ of n massless momenta $p_i$, i=1,...n, is spanned by $^{n}C_2$ number of Mandelstam variables,

\begin{equation}
s_{ij}=(p_i+p_j)^2=2p_i.p_j
\end{equation}

For space-time dimensions $d<n-1$, all of them are not linearly independent and they are constrained by
\begin{equation}
\Sigma_{j=1;j \neq i} s_{ij}=0,  ~~i=1,...,n
\end{equation}

Therefore the dimension of $\mathcal{K}_n$ reduces to
\begin{equation}
dim(\mathcal{K}_n)=^{n}C_2-n=\frac{n(n-3)}{2}
\end{equation}

For any set of particle labels
\begin{equation}
I \subset\{1,...,n\}
\end{equation}

one can define Mandelstamm variables as follows,

\begin{equation}
s_I=\left( \Sigma_{i \in I} p_i \right)^2=\Sigma_{i,j \in I; i<j}s_{ij}
\end{equation}

We always order the particles cyclically and define planar kinematic variables,
\begin{equation}
X_{i,j}=s_{i,i+1,...,j-1};~~1\leq i<j\leq n
\end{equation}

Here
\begin{equation}
s_{i,i+1,...j-1}=\left(p_i+p_{i+1}+...+p_{j-1}\right)^2
\end{equation}

which for massless particles is
\begin{equation}
2\left(p_i.p_{i+1}+...+p_{j-2}.p_{j-1}\right)
\end{equation}

Then $X_{i,i+1}=X_{1,n}=0$. The variables $X_{i,j}$ can be visualized as the diagonals between the $i^{th}$ and $j^{th}$ vertices of the corresponding n-gon.(See figure.)

\begin{figure}[H]
\centering
\definecolor{zzttqq}{rgb}{0.6,0.2,0}
\begin{tikzpicture}[line cap=round,line join=round,>=triangle 45,x=1cm,y=1cm]
%\clip(-7.66,-6.53) rectangle (13.66,6.53);
\fill[line width=2pt,color=zzttqq,fill=zzttqq,fill opacity=0.10000000149011612] (-2.68,-1.35) -- (-0.64,-1.31) -- (0.7742135623730948,0.16078210486801892) -- (0.7342135623730948,2.2007821048680185) -- (-0.7365685424949239,3.614995667241114) -- (-2.776568542494924,3.5749956672411143) -- (-4.190782104868019,2.1042135623730953) -- (-4.15078210486802,0.06421356237309528) -- cycle;
\draw [line width=2pt,color=zzttqq] (-2.68,-1.35)-- (-0.64,-1.31);
\draw [line width=2pt,color=zzttqq] (-0.64,-1.31)-- (0.7742135623730948,0.16078210486801892);
\draw [line width=2pt,color=zzttqq] (0.7742135623730948,0.16078210486801892)-- (0.7342135623730948,2.2007821048680185);
\draw [line width=2pt,color=zzttqq] (0.7342135623730948,2.2007821048680185)-- (-0.7365685424949239,3.614995667241114);
\draw [line width=2pt,color=zzttqq] (-0.7365685424949239,3.614995667241114)-- (-2.776568542494924,3.5749956672411143);
\draw [line width=2pt,color=zzttqq] (-2.776568542494924,3.5749956672411143)-- (-4.190782104868019,2.1042135623730953);
\draw [line width=2pt,color=zzttqq] (-4.190782104868019,2.1042135623730953)-- (-4.15078210486802,0.06421356237309528);
\draw [line width=2pt,color=zzttqq] (-4.15078210486802,0.06421356237309528)-- (-2.68,-1.35);
\draw [line width=2pt] (-4.190782104868019,2.1042135623730953)-- (-0.7365685424949239,3.614995667241114);
\draw [line width=2pt] (-4.190782104868019,2.1042135623730953)-- (0.7342135623730948,2.2007821048680185);
\draw [line width=2pt] (-4.15078210486802,0.06421356237309528)-- (0.7342135623730948,2.2007821048680185);
\draw [line width=2pt] (-2.68,-1.35)-- (0.7342135623730948,2.2007821048680185);
\draw [line width=2pt] (-2.68,-1.35)-- (0.7742135623730948,0.16078210486801892);
\draw (-4.84,0.19) node[anchor=north west] {1};
\draw (-4.9,2.39) node[anchor=north west] {2};
\draw (-2.98,4.31) node[anchor=north west] {3};
\draw (-0.6,4.19) node[anchor=north west] {4};
\draw (1,2.65) node[anchor=north west] {5};
\draw (0.98,0.25) node[anchor=north west] {6};
\draw (-0.58,-1.45) node[anchor=north west] {7};
\draw (-2.9,-1.71) node[anchor=north west] {8};
\draw (-3.04,3.45) node[anchor=north west] {$X_{24}$};
\draw (-0.92,2.81) node[anchor=north west] {$X_{25}$};
\draw (-3.36,1.27) node[anchor=north west] {$X_{15}$};
\draw (-2.64,0.07) node[anchor=north west] {$X_{58}$};
\draw (-0.36,0.55) node[anchor=north west] {$X_{68}$};
\end{tikzpicture}
\caption{Fully triangulated octagon. The $X_{ij}$ are the diagonals.
}
\end{figure}

These variables are related to the Mandelstam variables by the following relation
\begin{equation}
s_{ij}=X_{i,j+1}+X_{i+1,j}-X_{i,j}-X_{i+1,j+1}
\end{equation}

In other words $X_{i,j}$'s are dual to $\frac{n(n-3)}{2}$ diagonals of an n-gon made up of edges with momenta $p_1,...,p_n$. Each diagonal i.e., $X_{ij}$ cuts the internal propagator of a Feynman diagram once. Thus there exists a one-one correspondence between cuts of cubic graphs and complete triangulations of an n-gon.

A partial triangulation of a regular n-gon is a set of non-crossing diagonals which do not divide the n-gon into (n-2) triangles.

\definecolor{uuuuuu}{rgb}{0.26666666666666666,0.26666666666666666,0.26666666666666666}
\definecolor{zzttqq}{rgb}{0.6,0.2,0}
\definecolor{ududff}{rgb}{0.30196078431372547,0.30196078431372547,1}

\begin{figure}[H]
\begin{tikzpicture}[line cap=round,line join=round,>=triangle 45,x=1cm,y=1cm]
%\clip(1.5048716026882,-7.049036884832) rectangle (25.897017055726895,10.946879827084345);
\fill[line width=2pt,color=zzttqq,fill=zzttqq,fill opacity=0.10000000149011612] (5.887305636490103,-1.549254415893978) -- (10.447305636490103,-1.549254415893978) -- (11.856423130839861,2.7875632984119223) -- (8.167305636490102,5.467864048865599) -- (4.478188142140342,2.7875632984119223) -- cycle;
\fill[line width=2pt,color=zzttqq,fill=zzttqq,fill opacity=0.10000000149011612] (2.28,2.38) -- (3.34,2.36) -- (3.6865791443633467,3.361939567385364) -- (2.8407768353717517,4.001172274702884) -- (1.9714631162884584,3.3943002471603614) -- cycle;
\fill[line width=2pt,color=zzttqq,fill=zzttqq,fill opacity=0.10000000149011612] (2.76,-0.28) -- (3.84,-0.22) -- (4.116674962947234,0.8256820572612626) -- (3.2076694938847425,1.4119491100746369) -- (2.3691982550973476,0.7286000179362693) -- cycle;
\fill[line width=2pt,color=zzttqq,fill=zzttqq,fill opacity=0.10000000149011612] (3.68,-3.34) -- (4.94,-3.32) -- (5.310340282586531,-2.1154884495806074) -- (4.279223164628248,-1.39105937157959) -- (3.271617456761663,-2.1478491293556052) -- cycle;
\fill[line width=2pt,color=zzttqq,fill=zzttqq,fill opacity=0.10000000149011612] (7.6,-3.88) -- (8.78,-3.86) -- (9.125618923036534,-2.73157297088422) -- (8.159223164628246,-2.054166713066601) -- (7.216338816311659,-2.7639336506592183) -- cycle;
\fill[line width=2pt,color=zzttqq,fill=zzttqq,fill opacity=0.10000000149011612] (11.14,-3.42) -- (12.32,-3.46) -- (12.722682314014245,-2.350113990546718) -- (11.791553670743507,-1.624166713066601) -- (10.81340220728937,-2.2853926309967205) -- cycle;
\fill[line width=2pt,color=zzttqq,fill=zzttqq,fill opacity=0.10000000149011612] (12.16,-0.38) -- (13.14,-0.36) -- (13.423815524161546,0.5782157258567497) -- (12.619223164628247,1.1380649332158745) -- (11.838142215186648,0.5458550460817517) -- cycle;
\fill[line width=2pt,color=zzttqq,fill=zzttqq,fill opacity=0.10000000149011612] (12.58,2.4) -- (13.56,2.42) -- (13.843815524161545,3.3582157258567498) -- (13.039223164628247,3.9180649332158746) -- (12.258142215186648,3.3258550460817515) -- cycle;
\fill[line width=2pt,color=zzttqq,fill=zzttqq,fill opacity=0.10000000149011612] (10.42,4.48) -- (11.56,4.52) -- (11.874237112935637,5.616565108351472) -- (10.928446329256499,6.254279616189896) -- (10.029678365760756,5.551843748801479) -- cycle;
\fill[line width=2pt,color=zzttqq,fill=zzttqq,fill opacity=0.10000000149011612] (7.7,5.9) -- (8.78,5.92) -- (9.094717223599039,6.953321377486263) -- (8.209223164628249,7.571949110074636) -- (7.347240515749155,6.9209606977112665) -- cycle;
\fill[line width=2pt,color=zzttqq,fill=zzttqq,fill opacity=0.10000000149011612] (4.8,4.82) -- (5.9,4.8) -- (6.258939824138346,5.839981828037171) -- (5.380776835371752,6.502725945446391) -- (4.47910243651346,5.872342507812169) -- cycle;
\draw [line width=2pt,color=zzttqq] (5.887305636490103,-1.549254415893978)-- (10.447305636490103,-1.549254415893978);
\draw [line width=2pt,color=zzttqq] (10.447305636490103,-1.549254415893978)-- (11.856423130839861,2.7875632984119223);
\draw [line width=2pt,color=zzttqq] (11.856423130839861,2.7875632984119223)-- (8.167305636490102,5.467864048865599);
\draw [line width=2pt,color=zzttqq] (8.167305636490102,5.467864048865599)-- (4.478188142140342,2.7875632984119223);
\draw [line width=2pt,color=zzttqq] (4.478188142140342,2.7875632984119223)-- (5.887305636490103,-1.549254415893978);
\draw [line width=2pt,color=zzttqq] (2.28,2.38)-- (3.34,2.36);
\draw [line width=2pt,color=zzttqq] (3.34,2.36)-- (3.6865791443633467,3.361939567385364);
\draw [line width=2pt,color=zzttqq] (3.6865791443633467,3.361939567385364)-- (2.8407768353717517,4.001172274702884);
\draw [line width=2pt,color=zzttqq] (2.8407768353717517,4.001172274702884)-- (1.9714631162884584,3.3943002471603614);
\draw [line width=2pt,color=zzttqq] (1.9714631162884584,3.3943002471603614)-- (2.28,2.38);
\draw [line width=2pt] (1.9714631162884584,3.3943002471603614)-- (3.34,2.36);
\draw [line width=2pt] (1.9714631162884584,3.3943002471603614)-- (3.6865791443633467,3.361939567385364);
\draw [line width=2pt,color=zzttqq] (2.76,-0.28)-- (3.84,-0.22);
\draw [line width=2pt,color=zzttqq] (3.84,-0.22)-- (4.116674962947234,0.8256820572612626);
\draw [line width=2pt,color=zzttqq] (4.116674962947234,0.8256820572612626)-- (3.2076694938847425,1.4119491100746369);
\draw [line width=2pt,color=zzttqq] (3.2076694938847425,1.4119491100746369)-- (2.3691982550973476,0.7286000179362693);
\draw [line width=2pt,color=zzttqq] (2.3691982550973476,0.7286000179362693)-- (2.76,-0.28);
\draw [line width=2pt,color=zzttqq] (3.68,-3.34)-- (4.94,-3.32);
\draw [line width=2pt,color=zzttqq] (4.94,-3.32)-- (5.310340282586531,-2.1154884495806074);
\draw [line width=2pt,color=zzttqq] (5.310340282586531,-2.1154884495806074)-- (4.279223164628248,-1.39105937157959);
\draw [line width=2pt,color=zzttqq] (4.279223164628248,-1.39105937157959)-- (3.271617456761663,-2.1478491293556052);
\draw [line width=2pt,color=zzttqq] (3.271617456761663,-2.1478491293556052)-- (3.68,-3.34);
\draw [line width=2pt,color=zzttqq] (7.6,-3.88)-- (8.78,-3.86);
\draw [line width=2pt,color=zzttqq] (8.78,-3.86)-- (9.125618923036534,-2.73157297088422);
\draw [line width=2pt,color=zzttqq] (9.125618923036534,-2.73157297088422)-- (8.159223164628246,-2.054166713066601);
\draw [line width=2pt,color=zzttqq] (8.159223164628246,-2.054166713066601)-- (7.216338816311659,-2.7639336506592183);
\draw [line width=2pt,color=zzttqq] (7.216338816311659,-2.7639336506592183)-- (7.6,-3.88);
\draw [line width=2pt,color=zzttqq] (11.14,-3.42)-- (12.32,-3.46);
\draw [line width=2pt,color=zzttqq] (12.32,-3.46)-- (12.722682314014245,-2.350113990546718);
\draw [line width=2pt,color=zzttqq] (12.722682314014245,-2.350113990546718)-- (11.791553670743507,-1.624166713066601);
\draw [line width=2pt,color=zzttqq] (11.791553670743507,-1.624166713066601)-- (10.81340220728937,-2.2853926309967205);
\draw [line width=2pt,color=zzttqq] (10.81340220728937,-2.2853926309967205)-- (11.14,-3.42);
\draw [line width=2pt,color=zzttqq] (12.16,-0.38)-- (13.14,-0.36);
\draw [line width=2pt,color=zzttqq] (13.14,-0.36)-- (13.423815524161546,0.5782157258567497);
\draw [line width=2pt,color=zzttqq] (13.423815524161546,0.5782157258567497)-- (12.619223164628247,1.1380649332158745);
\draw [line width=2pt,color=zzttqq] (12.619223164628247,1.1380649332158745)-- (11.838142215186648,0.5458550460817517);
\draw [line width=2pt,color=zzttqq] (11.838142215186648,0.5458550460817517)-- (12.16,-0.38);
\draw [line width=2pt,color=zzttqq] (12.58,2.4)-- (13.56,2.42);
\draw [line width=2pt,color=zzttqq] (13.56,2.42)-- (13.843815524161545,3.3582157258567498);
\draw [line width=2pt,color=zzttqq] (13.843815524161545,3.3582157258567498)-- (13.039223164628247,3.9180649332158746);
\draw [line width=2pt,color=zzttqq] (13.039223164628247,3.9180649332158746)-- (12.258142215186648,3.3258550460817515);
\draw [line width=2pt,color=zzttqq] (12.258142215186648,3.3258550460817515)-- (12.58,2.4);
\draw [line width=2pt,color=zzttqq] (10.42,4.48)-- (11.56,4.52);
\draw [line width=2pt,color=zzttqq] (11.56,4.52)-- (11.874237112935637,5.616565108351472);
\draw [line width=2pt,color=zzttqq] (11.874237112935637,5.616565108351472)-- (10.928446329256499,6.254279616189896);
\draw [line width=2pt,color=zzttqq] (10.928446329256499,6.254279616189896)-- (10.029678365760756,5.551843748801479);
\draw [line width=2pt,color=zzttqq] (10.029678365760756,5.551843748801479)-- (10.42,4.48);
\draw [line width=2pt,color=zzttqq] (7.7,5.9)-- (8.78,5.92);
\draw [line width=2pt,color=zzttqq] (8.78,5.92)-- (9.094717223599039,6.953321377486263);
\draw [line width=2pt,color=zzttqq] (9.094717223599039,6.953321377486263)-- (8.209223164628249,7.571949110074636);
\draw [line width=2pt,color=zzttqq] (8.209223164628249,7.571949110074636)-- (7.347240515749155,6.9209606977112665);
\draw [line width=2pt,color=zzttqq] (7.347240515749155,6.9209606977112665)-- (7.7,5.9);
\draw [line width=2pt,color=zzttqq] (4.8,4.82)-- (5.9,4.8);
\draw [line width=2pt,color=zzttqq] (5.9,4.8)-- (6.258939824138346,5.839981828037171);
\draw [line width=2pt,color=zzttqq] (6.258939824138346,5.839981828037171)-- (5.380776835371752,6.502725945446391);
\draw [line width=2pt,color=zzttqq] (5.380776835371752,6.502725945446391)-- (4.47910243651346,5.872342507812169);
\draw [line width=2pt,color=zzttqq] (4.47910243651346,5.872342507812169)-- (4.8,4.82);
\draw [line width=2pt] (2.3691982550973476,0.7286000179362693)-- (3.84,-0.22);
\draw [line width=2pt] (3.271617456761663,-2.1478491293556052)-- (4.94,-3.32);
\draw [line width=2pt] (4.279223164628248,-1.39105937157959)-- (4.94,-3.32);
\draw [line width=2pt] (8.159223164628246,-2.054166713066601)-- (8.78,-3.86);
\draw [line width=2pt] (11.791553670743507,-1.624166713066601)-- (12.32,-3.46);
\draw [line width=2pt] (11.791553670743507,-1.624166713066601)-- (11.14,-3.42);
\draw [line width=2pt] (12.619223164628247,1.1380649332158745)-- (12.16,-0.38);
\draw [line width=2pt] (13.039223164628247,3.9180649332158746)-- (12.58,2.4);
\draw [line width=2pt] (12.58,2.4)-- (13.843815524161545,3.3582157258567498);
\draw [line width=2pt] (11.874237112935637,5.616565108351472)-- (10.42,4.48);
\draw [line width=2pt] (9.094717223599039,6.953321377486263)-- (7.7,5.9);
\draw [line width=2pt] (9.094717223599039,6.953321377486263)-- (7.347240515749155,6.9209606977112665);
\draw [line width=2pt] (6.258939824138346,5.839981828037171)-- (4.47910243651346,5.872342507812169);
\begin{scriptsize}
\draw [fill=ududff] (5.887305636490103,-1.549254415893978) circle (2.5pt);
\draw [fill=ududff] (10.447305636490103,-1.549254415893978) circle (2.5pt);
\draw [fill=uuuuuu] (11.856423130839861,2.7875632984119223) circle (2.5pt);
\draw [fill=uuuuuu] (8.167305636490102,5.467864048865599) circle (2.5pt);
\draw [fill=uuuuuu] (4.478188142140342,2.7875632984119223) circle (2.5pt);
\draw [fill=ududff] (2.28,2.38) circle (2.5pt);
\draw [fill=ududff] (3.34,2.36) circle (2.5pt);
\draw [fill=uuuuuu] (3.6865791443633467,3.361939567385364) circle (2.5pt);
\draw [fill=uuuuuu] (2.8407768353717517,4.001172274702884) circle (2.5pt);
\draw [fill=uuuuuu] (1.9714631162884584,3.3943002471603614) circle (2.5pt);
\draw [fill=ududff] (2.76,-0.28) circle (2.5pt);
\draw [fill=ududff] (3.84,-0.22) circle (2.5pt);
\draw [fill=uuuuuu] (4.116674962947234,0.8256820572612626) circle (2.5pt);
\draw [fill=uuuuuu] (3.2076694938847425,1.4119491100746369) circle (2.5pt);
\draw [fill=uuuuuu] (2.3691982550973476,0.7286000179362693) circle (2.5pt);
\draw [fill=ududff] (3.68,-3.34) circle (2.5pt);
\draw [fill=ududff] (4.94,-3.32) circle (2.5pt);
\draw [fill=uuuuuu] (5.310340282586531,-2.1154884495806074) circle (2.5pt);
\draw [fill=uuuuuu] (4.279223164628248,-1.39105937157959) circle (2.5pt);
\draw [fill=uuuuuu] (3.271617456761663,-2.1478491293556052) circle (2.5pt);
\draw [fill=ududff] (7.6,-3.88) circle (2.5pt);
\draw [fill=ududff] (8.78,-3.86) circle (2.5pt);
\draw [fill=uuuuuu] (9.125618923036534,-2.73157297088422) circle (2.5pt);
\draw [fill=uuuuuu] (8.159223164628246,-2.054166713066601) circle (2.5pt);
\draw [fill=uuuuuu] (7.216338816311659,-2.7639336506592183) circle (2.5pt);
\draw [fill=ududff] (11.14,-3.42) circle (2.5pt);
\draw [fill=ududff] (12.32,-3.46) circle (2.5pt);
\draw [fill=uuuuuu] (12.722682314014245,-2.350113990546718) circle (2.5pt);
\draw [fill=uuuuuu] (11.791553670743507,-1.624166713066601) circle (2.5pt);
\draw [fill=uuuuuu] (10.81340220728937,-2.2853926309967205) circle (2.5pt);
\draw [fill=ududff] (12.16,-0.38) circle (2.5pt);
\draw [fill=ududff] (13.14,-0.36) circle (2.5pt);
\draw [fill=uuuuuu] (13.423815524161546,0.5782157258567497) circle (2.5pt);
\draw [fill=uuuuuu] (12.619223164628247,1.1380649332158745) circle (2.5pt);
\draw [fill=uuuuuu] (11.838142215186648,0.5458550460817517) circle (2.5pt);
\draw [fill=ududff] (12.58,2.4) circle (2.5pt);
\draw [fill=ududff] (13.56,2.42) circle (2.5pt);
\draw [fill=uuuuuu] (13.843815524161545,3.3582157258567498) circle (2.5pt);
\draw [fill=uuuuuu] (13.039223164628247,3.9180649332158746) circle (2.5pt);
\draw [fill=uuuuuu] (12.258142215186648,3.3258550460817515) circle (2.5pt);
\draw [fill=ududff] (10.42,4.48) circle (2.5pt);
\draw [fill=ududff] (11.56,4.52) circle (2.5pt);
\draw [fill=uuuuuu] (11.874237112935637,5.616565108351472) circle (2.5pt);
\draw [fill=uuuuuu] (10.928446329256499,6.254279616189896) circle (2.5pt);
\draw [fill=uuuuuu] (10.029678365760756,5.551843748801479) circle (2.5pt);
\draw [fill=ududff] (7.7,5.9) circle (2.5pt);
\draw [fill=ududff] (8.78,5.92) circle (2.5pt);
\draw [fill=uuuuuu] (9.094717223599039,6.953321377486263) circle (2.5pt);
\draw [fill=uuuuuu] (8.209223164628249,7.571949110074636) circle (2.5pt);
\draw [fill=uuuuuu] (7.347240515749155,6.9209606977112665) circle (2.5pt);
\draw [fill=ududff] (4.8,4.82) circle (2.5pt);
\draw [fill=ududff] (5.9,4.8) circle (2.5pt);
\draw [fill=uuuuuu] (6.258939824138346,5.839981828037171) circle (2.5pt);
\draw [fill=uuuuuu] (5.380776835371752,6.502725945446391) circle (2.5pt);
\draw [fill=uuuuuu] (4.47910243651346,5.872342507812169) circle (2.5pt);
\end{scriptsize}
\end{tikzpicture}

\caption{The 2-dimensional associahedron $\mathcal{A}_5$ 5 partial triangulations represented by 5 diagonals on co-dim 1 faces.5 complete triangulations represented by 5 vertices on co-dim 0 vertices.}
\end{figure}

The associahedron of dimension (n-3) is a polytope whose co-dimension $d$ boundaries are in one-to-one correspondence with partial triangulations by $d$ diagonals. The vertices represent complete triangulations and the $k$-faces represent $k$ partial triangulations of the n-gon.
The total number of ways to triangulate a convex n-gon by non-intersecting diagonals is the (n-2)-eth Catalan number, $C_{n-2}=\frac{1}{n-1} ~~^{2n-4}C_{n-2}$.

\begin{figure}[H]
\centering
\includegraphics[trim={0 0cm 0 0cm}, clip,width=\linewidth]{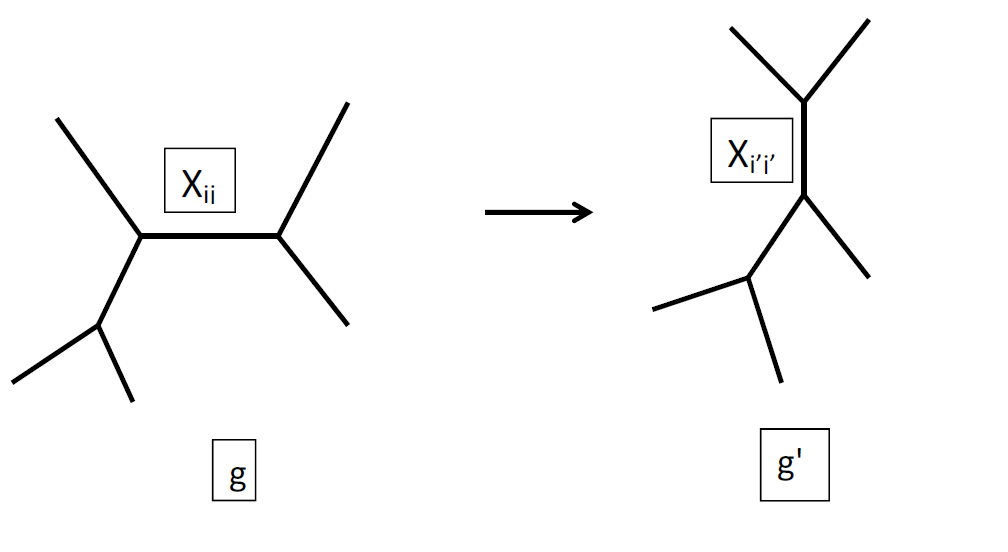}
\caption{Two graphs related by a mutation given by an exchange of $X_{ij} \to X_{i'j'}$ in a 4-point sub-graph.}\label{mutationdiag}

\end{figure}

We now define the planar scattering form. It is a differential form on the space of kinematic variables $X_{ij}$ that encodes information about on-shell tree-level scattering amplitudes of the scalar $\phi^3$ theory. Let g denote a tree cubic graph with propagators $X_{i_aj_a}$ for $a=1,...,n-3$. For each ordering of these propagators, we assigns a value $sign(g)\in~{\pm}$ to the graph with the property that flipping two propagators flips the sign. The form must have logarithmic singularities at $X_{i_aj_a}=0$. Therefore one assigns to the graph a d log form and thus
defines the planar scattering form of rank (n-3) :

\begin{equation}\label{scatteringform}
\Omega^{n-3}_n:=\Sigma_{planar~g}~ sign(g) \wedge^{n-3}_{a=1} dlog X_{i_aj_a}
\end{equation}

where the sum is over each planar cubic graph g. There two sign choices for each graph, therefore there are many different scattering forms. One can fix the scattering form uniquely if one demands projectivity of the differential form, i.e. if one requires that the form should be invariant under local $GL(1)$ transformations $X_{ij}\to \Lambda(X)X_{ij}$ for any index pair (i,j). We use projectivity to define an operation called mutation. Two planar sub-graphs g and g' are related by a mutation if we can obtain one from the other just by exchanging 4-point sub-graph channels, figure \ref{mutationdiag}. As we can see $X_{ij}$ and $X_{i'j'}$ are the mutated propagators of graphs g and g' respectively. Let's denote the rest of the common propagators as $X_{i_bj_b}$ with $b=1,...,n-4$. Under $X_{ij}\to \Lambda(X)X_{ij}$ the scattering form becomes,

\begin{equation}
sign(g) ~ dlog(\Lambda(X)X_{ij})\Lambda_{b=1}^{n-4}~dlogX_{i_bj_b}  ~~+...
\end{equation}

\begin{equation}
sign(g') ~ dlog(\Lambda(X)X_{i'j'})\Lambda_{b=1}^{n-4}~dlogX_{i_bj_b}  ~~+...
\end{equation}

Collecting the $dlog\Lambda$ terms

\begin{equation}
sign(g) ~ dlog\Lambda(X)\Lambda_{b=1}^{n-4}~dlogX_{i_bj_b} \\ +sign(g') ~ dlog\Lambda(X)\Lambda_{b=1}^{n-4}~dlogX_{i_bj_b}  
\end{equation}

\begin{equation}
=(sign(g)+sign(g')) ~ dlog(\Lambda(X)X_{ij}\Lambda_{b=1}^{n-4}~dlogX_{i_bj_b}  
\end{equation}

Since we demand projectivity, the $\Lambda(X)$ dependence has to disappear, i.e. when
\begin{equation}
sign(g)=-sign(g')
\end{equation}

for each mutation.

\subsection{The kinematic associahedron.}

We described above how one gets an associahedron $\mathcal{A}_n$ inside the kinematic space $\mathcal{K}_n$ but it is not evident how it should be embedded in $\mathcal{K}_n$. Because $\mathcal{A}_n$ and $\mathcal{K}_n$ are of different dimensionality, $dim(\mathcal{A}_n)=(n-3)$ and $dim(\mathcal{K}_n)=\frac{n(n-3)}{2}$ respectively, we have to impose constraints to embed $\mathcal{A}_n$ inside $\mathcal{K}_n$. One natural choice is to demand all planar kinematic variables being positive,

\begin{equation}
X_{ij}\geq;~~1\leq i<j\leq n.
\end{equation}

These are $\frac{n(n-3)}{2}$ inequalities and thus cut out a big simplex $\Delta_n$ inside $\mathcal{K}_n$ which is still $\frac{n(n-3)}{2}$ dimensional. Therefore we need $\frac{n(n-3)}{2}-n-3=\frac{(n-2)(n-3)}{2}$ more constraints to embed $\mathcal{A}_n$ inside $\mathcal{K}_n$. To do that we impose the following constraints,

\begin{equation}
s_{ij}=-c_{ij};   ~~~~1\leq i<j \leq n-1, ~~|i-j|\geq 2
\end{equation}

where $c_{ij}$ are positive constraints.\\

       These constraints give a space $\mathcal{H}_n$ of dimension (n-3) which is precisely the dimension of $\mathcal{A}_n$. The kinematic associahedron $\mathcal{A}_n$ can now be embedded in $\mathcal{K}_n$ as the intersection of the simplex $\Delta_n$ and the subspace $\mathcal{H}_n$ as follows,
       
\begin{equation}
\mathcal{A}_n:=\mathcal{H}_n\cap \Delta_n.
\end{equation}

Once one has the associahedron in $\mathcal{K}_n$ all one has to do is to obtain its canonical form $\Omega(\mathcal{A}_n)$.

Since an associahedron is a simple polytope one can directly write down its canonical form as follows,

\begin{equation}\label{canonicalform}
\Omega(\mathcal{A}_n)=\Sigma_{vertex~Z}~sign(Z)\Lambda_{a=1}^{n-3}dlogX_{i_aj_a}
\end{equation}

 Here for each vertex Z, $X_{i_aj_a}=0$ denote its adjacent facets for a=1,...,n-3. Now we claim that the above differential form \ref{canonicalform} is identical to the pullback of the scattering form \ref{scatteringform} in $\mathcal{K}_n$ to the subspace $\mathcal{H}_n$. We can justify this statement by the identification $g\leftrightarrow Z$ and $sign(g)\leftrightarrow sign(Z)$.

\begin{figure}[H]
\centering
\includegraphics[trim={0 0cm 0 0cm}, clip,width=\linewidth]{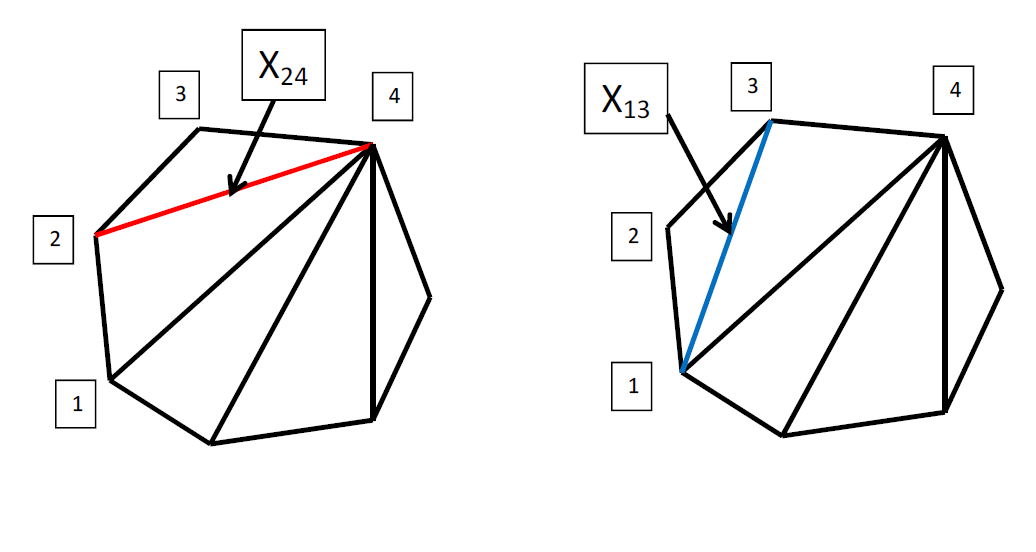}
\caption{Two triangles related by a mutation $X_{24}\to X_{13}$.}\label{triangmut}

\end{figure}

\begin{enumerate}

\item
There is a one to one correspondence between vertices Z and the planar cubic graphs g. Also, g and its corresponding vertex have the same propagators $X_{i_aj_a}$.

\item
Let Z and Z' be two vertices related by mutation. Note that mutation can also be framed in the language of triangulation. Two triangulations are related by a mutation if one can be obtained from the other by exchanging exactly one diagonal. See figure \ref{triangmut}.\\

Thus for Z and Z' vertices we have

\begin{equation}
\Lambda_{a=1}^{n-3}dX_{i_aj_a}=-\Lambda_{a=1}^{n-3}dX_{i'_aj'_a}
\end{equation}

which leads to the sign-flip rule $sign(Z)=-sign(Z')$.

\end{enumerate}

Therefore one can construct the following quantity (an (n-3)-form) which is independent of g on pullback.

\begin{equation}
d^{n-3}X:=sign(g)\Lambda_{a=1}^{n-3}dX_{i_aj_a}.
\end{equation}

Substituting this in \ref{canonicalform} one gets,

\begin{equation}
\Omega(\mathcal{A}_n= \Sigma_{planar ~g} ~~\frac{1}{\Pi_{a=1}^{n-3} X_{i_aj_a}} ~~d^{n-3}X.
\end{equation}

Here

\begin{equation}
\mathcal{M}_n=\Sigma_{planar ~g} ~~\frac{1}{\Pi_{a=1}^{n-3} X_{i_aj_a}}
\end{equation}

is the tree level n-point scattering amplitude for the cubic scalar theory.

\section{Known results of quivers for tree level and 1-loop processes}

In this section we review known results for tree level and 1-loop quivers.

\subsection{Tree level}

\begin{figure}[H]
\centering
\includegraphics[trim={0 1cm 0 4.5cm}, clip,width=\linewidth]{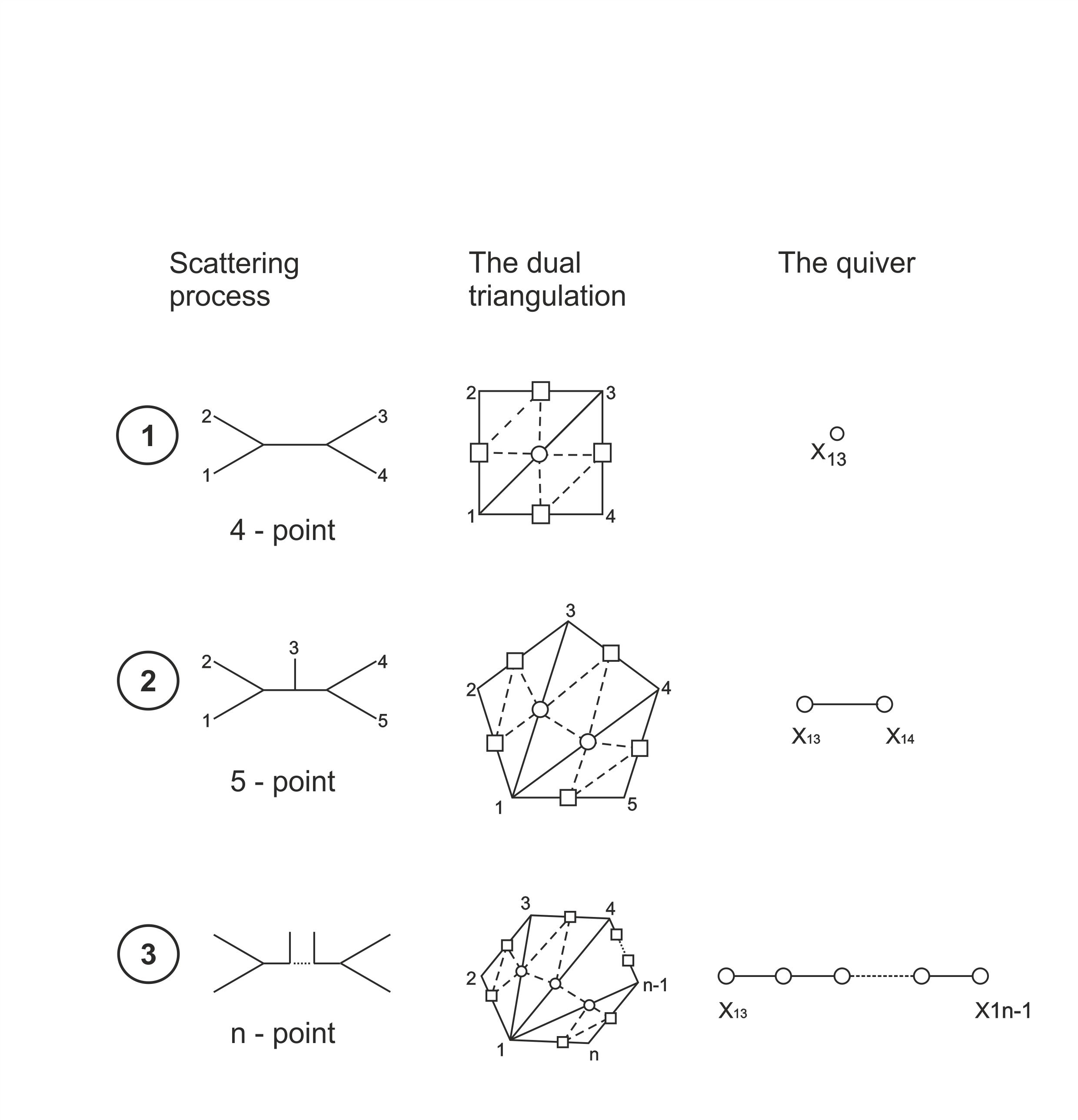}
\caption{Demonstration of a few examples of quivers at tree level.}\label{treelevelmod}

\end{figure}

The quiver for the $4$-point tree level scattering amplitude is given by a single node. As the number of particles increases, the number of internal lines also increases. The number of nodes of the quiver is determined by the number of internal lines. Thus for a 5-point scattering process, there are two internal lines, therefore two nodes, as shown in figure \ref{treelevelmod}. Therefore for an $n$-point amplitude, there are $n-1$ nodes in the quiver. Also, in the figure are the triangulations of the dual $n$-gons whose diagonals correspond to the internal lines and the quiver nodes. To get quivers from the triangulated diagrams, one associates circle shaped nodes to the diagonals,as shown. These are called unfrozen nodes. Also, we associate square shaped nodes to the external polygon. These are called frozen nodes. Now one connects all nodes without crossing any diagonals. This gives us the quiver for the particular triangulation.

In part 1 of figure \ref{treelevelmod}, there are four frozen nodes and one unfrozen node. Thus we associate to this triangulation an $A_1$ type of Dynkin quiver. In part 2, there are two unfrozen nodes and connecting these gives us the $A_2$ type Dynkin quiver. In part 3, we have shown an $A_{n-1}$ quiver, which corresponds to the $n-1$ diagonals/unfrozen nodes of the triangulations.\\

\subsection{All unique 1-loop processes}\label{All unique 1-loop processes}

\begin{figure}[H]
\centering
\includegraphics[trim={0 1cm 0 10cm 0},clip,width=\linewidth]{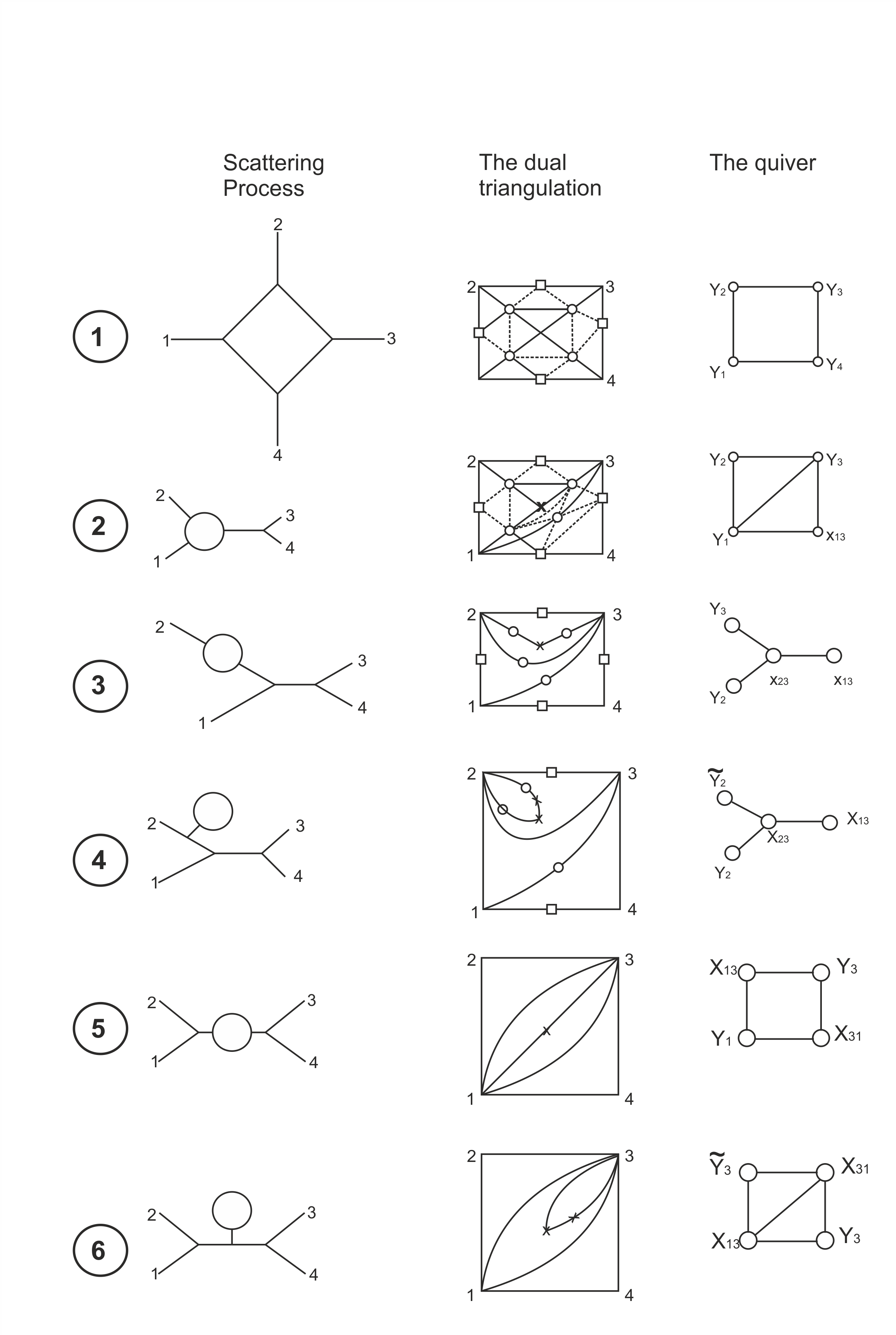}
\caption{Quivers for $\phi^3$ theories at 1-loop.}\label{1loopmod}
\end{figure}

In figure \ref{1loopmod} we have tabulated 4-particle scattering processes with their dual n-gons(in this case, squares) triangulated and the associated quivers. We have only shown those processes which are unique upto permutations. Again as in the case of tree level diagrams one has to triangulate and mark the frozen and unfrozen nodes and connect them. An additional structure that is added to the polygon at 1-loop is the marked point in the middle of the diagram. This corresponds to the singularity of the loop. Thus for the case of 4-particles, we have a 4-gon, a square, with a marked point, and this has to be triangulated. This gives us four triangles with four unfrozen nodes and 4 diagonals. In part 1 and 2 we have shown the box diagram and the vertex corrections with their triangulations. We have also marked the frozen and the unfrozen nodes and connected them in the middle picture. Then we have shown the quiver by drawing the connected unfrozen nodes in the rightmost picture. The quivers, as can be seen, in this case are a square and a square with a diagonal. This is already a deviation from the association of $D_4$ Dynkin diagrams with the quivers. In parts 3 and 4 we have drawn the processes and only marked the frozen and unfrozen nodes. Then we have shown the quiver associated to it. It is however seen that it is only in these two cases that one gets to see the $D_4$ Dynkin diagrams.
 
 In parts 5 and 6 we only show the processes, triangulations and the associated quivers. One can again see that the quivers in this case are a square and a square with a diagonal, not $D_n$ type.

\section{Construction of quivers from Feynman diagrams}

We will discuss the rules for constructing quivers for $\phi^3$ theory. These can be used to construct quivers at all loop orders at all particle numbers. We will also see examples in the following subsections upto 3-loops.

\subsection{Feynman diagram-like rules for writing down quivers from Feynman diagrams}

\begin{figure}[H]
\centering
\includegraphics[trim={0 0cm 0 1cm 0},clip,width=\linewidth]{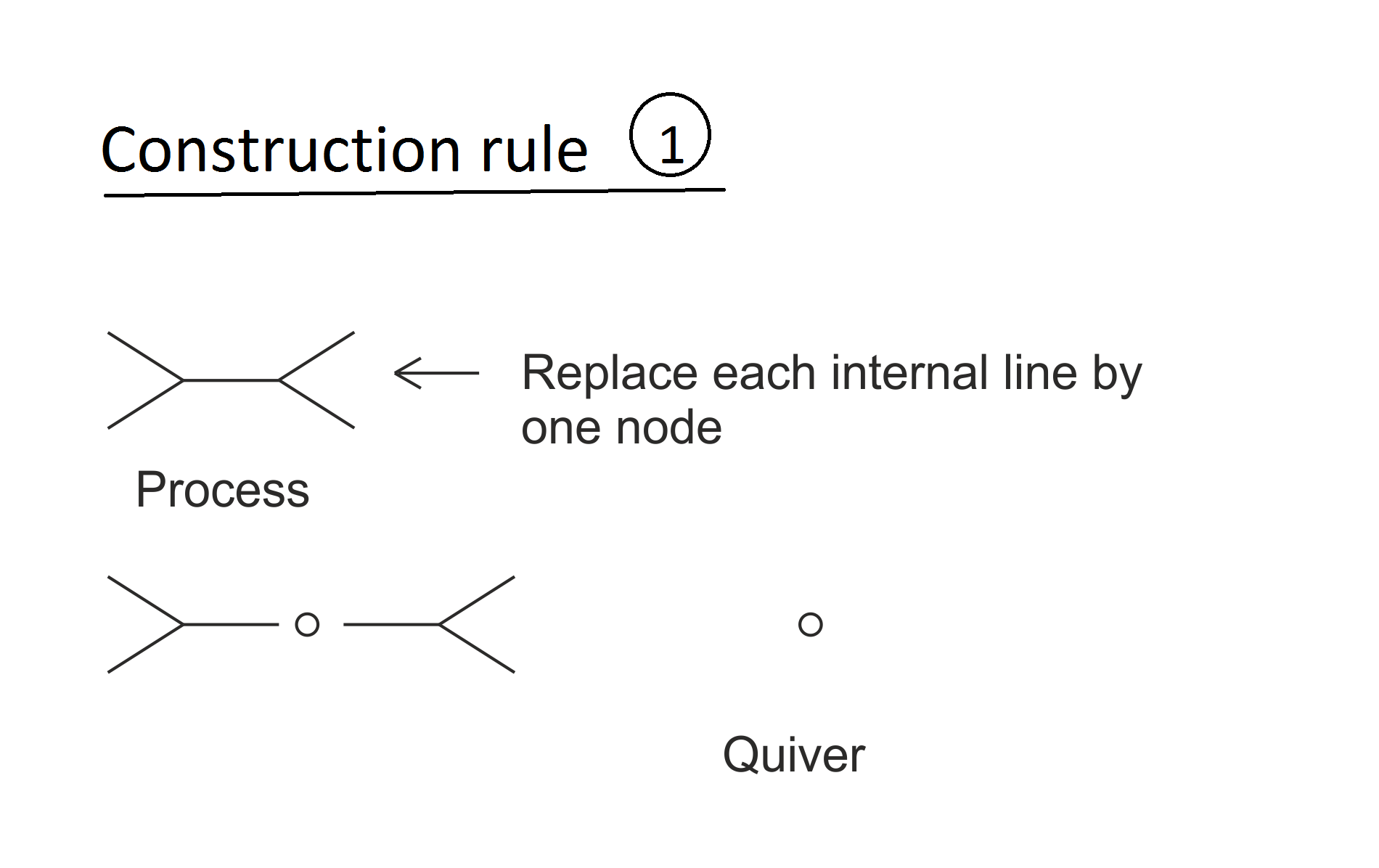}
\caption{Construction rule 1.}\label{construles1}
\end{figure}

\begin{figure}[H]
\centering
\includegraphics[trim={0 0cm 0 0cm 0},clip,width=\linewidth]{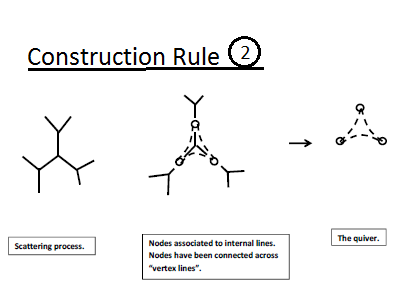}
\caption{Construction rule 2.}\label{construles2}
\end{figure}

Rules:

\begin{enumerate}
\item
Replace each internal line by a node as shown in the figure \ref{construles1}. Therefore the quiver of a 4-point scattering process is an $A_1$ quiver.
\item
Associate a node to each internal line of a $\phi^3$ vertex, then you connect two nodes along each "vertex line" as shown in the figure. Thus three nodes associated to the three lines of the $\phi^3$ vertex have an associated quiver that is a triangle. (Ref. fig. \ref{construles2}).
\end{enumerate}

We will append more rules to the list along the way.\\

\begin{figure}[H]
\centering
\includegraphics[trim={0 1cm 0 0cm 0},clip,width=\linewidth]{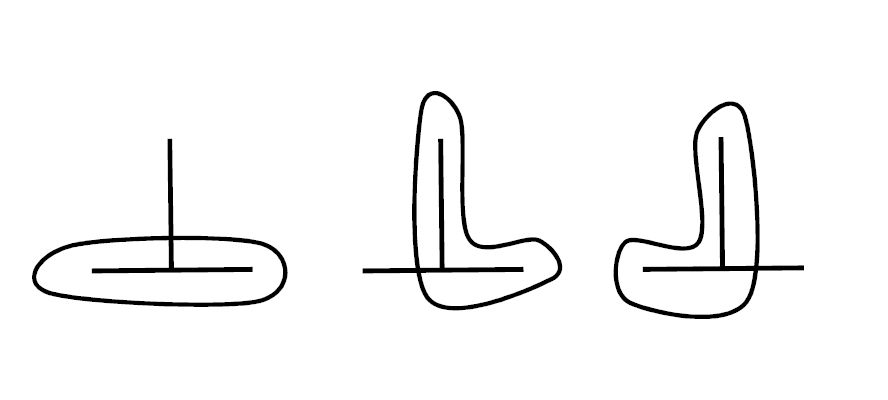}
\caption{The three "vertex lines" of the $\phi^3$ vertex.}\label{vertexline}
\end{figure}

\subsection{Tree level diagrams}
At the tree level we have already seen the type of quivers one gets as one goes higher in particle numbers above, Fig. \ref{treelevelmod}. We will show the construction of the quivers for the tree level processes for the 4,5 and the n particle cases, as we did above, using our technique.

For the 4 point case, there is a single internal line and therefore we get an $A_1$ quiver. For the 5-point there are two internal lines connected by one $\phi^3$ vertex. We join the nodes across the "vertex lines". We get the $A_2$ quiver. For an n-point case we have $n-1$ nodes connected by n-2 lines. We connect all nodes by vertex lines and get the $A_{n-1}$ quiver.\\

\begin{figure}[H]
\centering
\includegraphics[trim={0 1cm 0 1.5cm 0},clip,width=\linewidth]{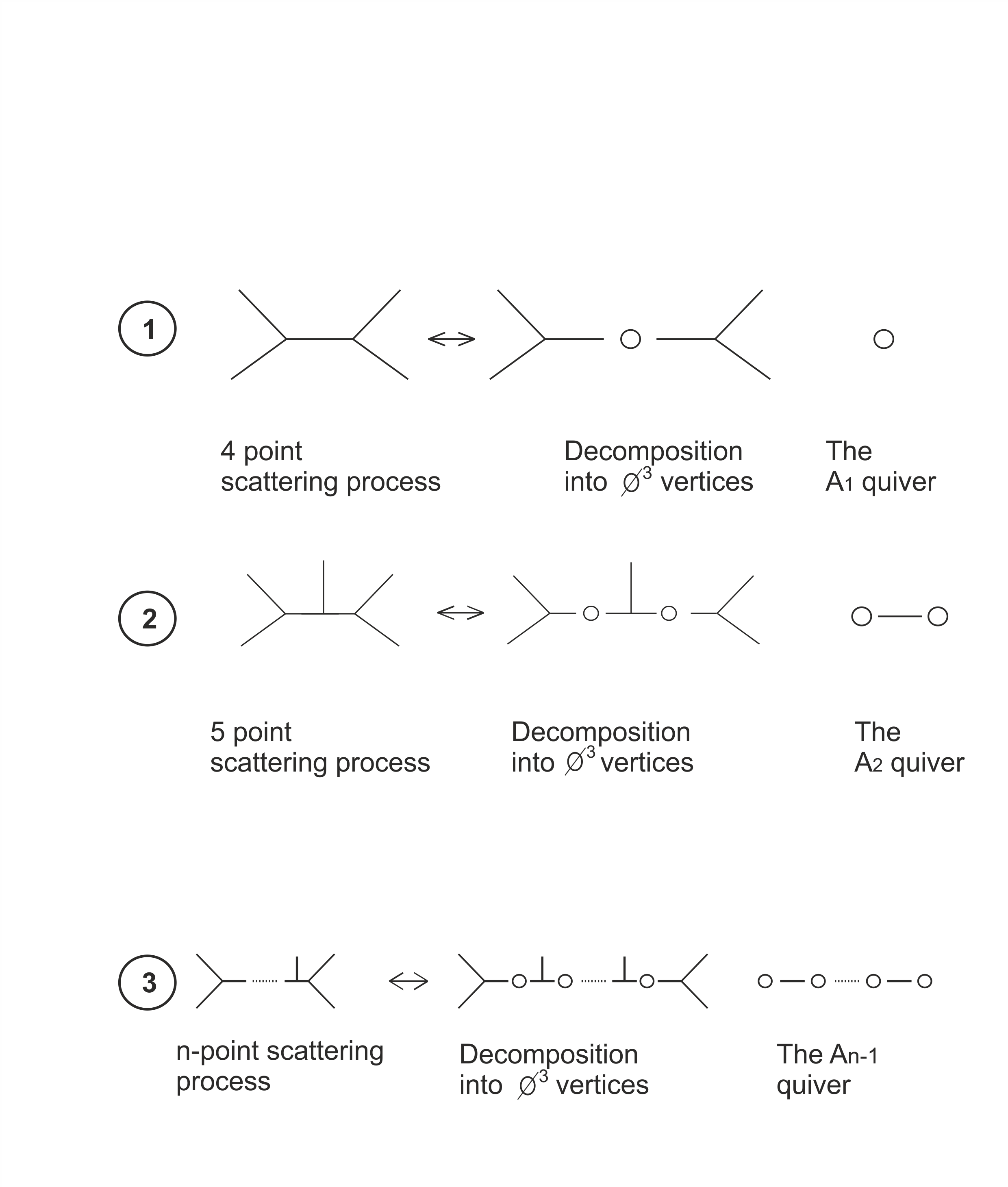}
\caption{Construction of quivers for tree level processes.}\label{treequivmod}
\end{figure}

\subsection{1-loop diagrams}\label{1loopquivers}
Here we show the construction of the quivers for the 1-loop diagrams. In part 1 of the figure \ref{1loopquiv1mod}, we have the box diagram for a 4 particle scattering. In the next figure we decompose the diagram into its 3-point vertices as shown on the top right diagram. We associate nodes to all internal lines and connect the nodes across all "vertex lines". We do the same thing for diagram 2 in the figure. One can see in the bottom right diagram that there is an additional connection between nodes $Y_1$ and $Y_3$, which is exactly what we saw in the construction of the quiver from the earlier method.\\

\begin{figure}[H]
\centering
\includegraphics[trim={0 1cm 0 1.5cm 0},clip,width=\linewidth]{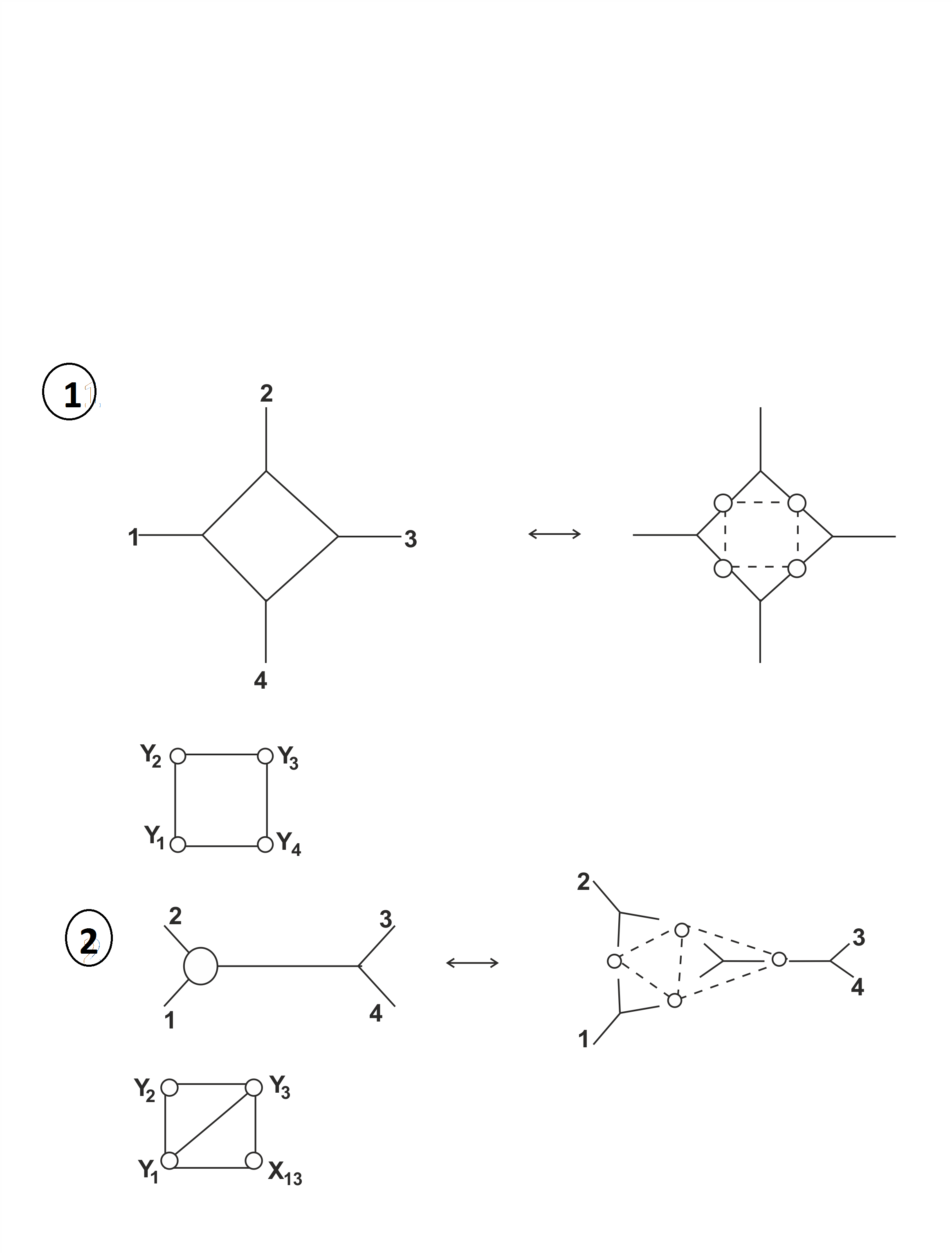}
\caption{Construction of quivers for 4-point scattering processes at 1-loop}\label{1loopquiv1mod}
\end{figure}

\begin{figure}[H]
\centering
\includegraphics[trim={0 1cm 0 1.5cm 0},clip,width=\linewidth]{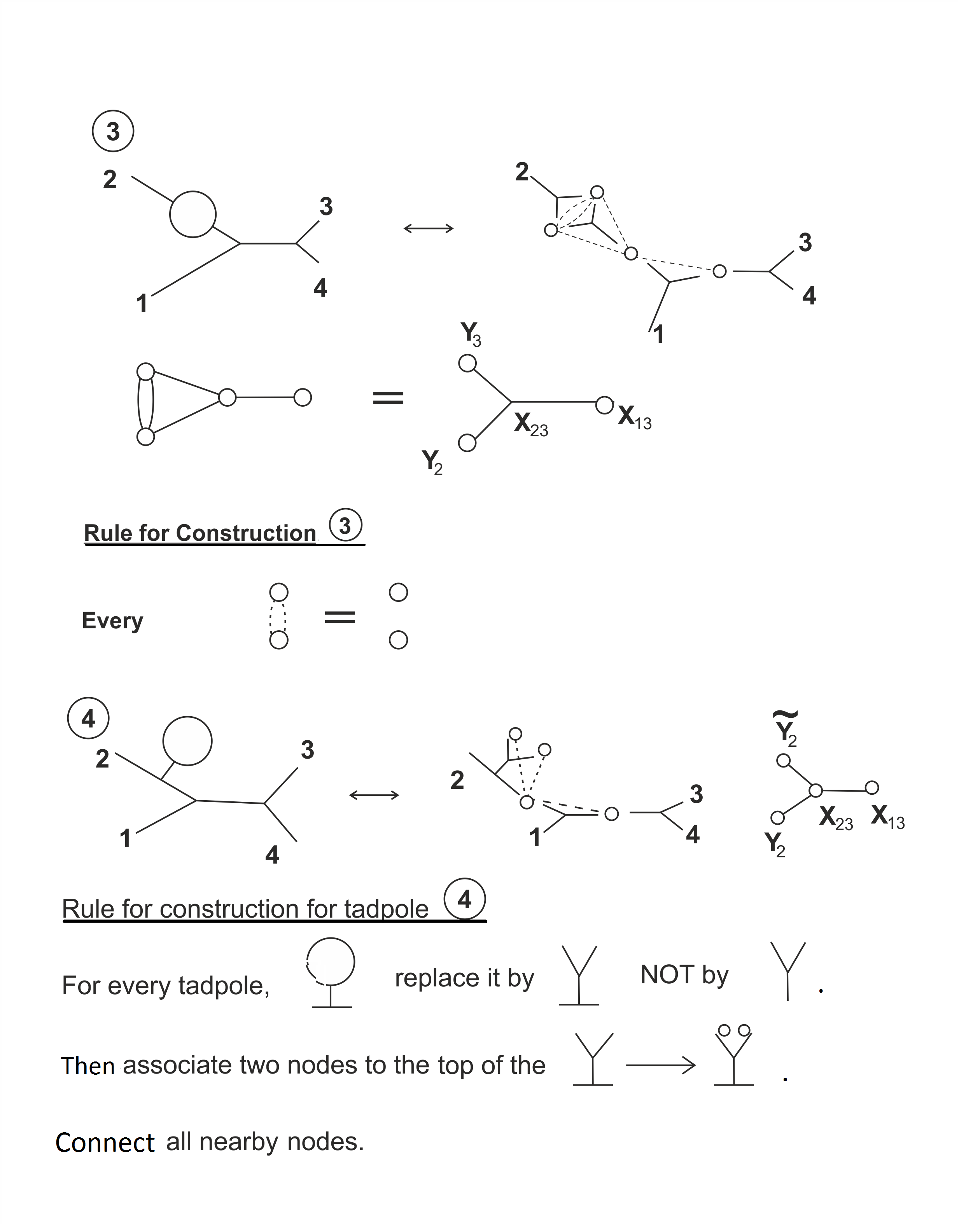}
\caption{Construction rules 3 and 4.}\label{1loopquiv2mod}
\end{figure}

In part 3 of figure \ref{1loopquiv2mod}, we can see that we come across a circular construction in one part of the quiver. This leads us to another rule,\\

Rule 3: Closed loops in quivers cancel out.\\

In part 4 of the figure we have a tadpole. We need another rule for this. \\

\hypertarget{$\phi^3$ rule 4.}{Rule 4:}
\\
For each tadpole with its base, replace it by a "Y" with its base, i.e. separate the tadpole from the rest of the diagram and cut its loop so that it becomes a "Y". Then associate two nodes to each of the open branches(two) of the "Y". Then we connect all nearby nodes with both these nodes.\\

\iffalse

From process 4 we see that another rule has to be added for constructing quivers from diagrams with tadpoles.\\
\\
Rule 4: Do a special replacement for the tadpole as shown. Then associate two nodes to it as shown and connect all nearby nodes. We can see this for the construction of quiver for process 4.

\fi

\begin{figure}[H]
\centering
\includegraphics[trim={0 1cm 0 1.5cm 0},clip,width=\linewidth]{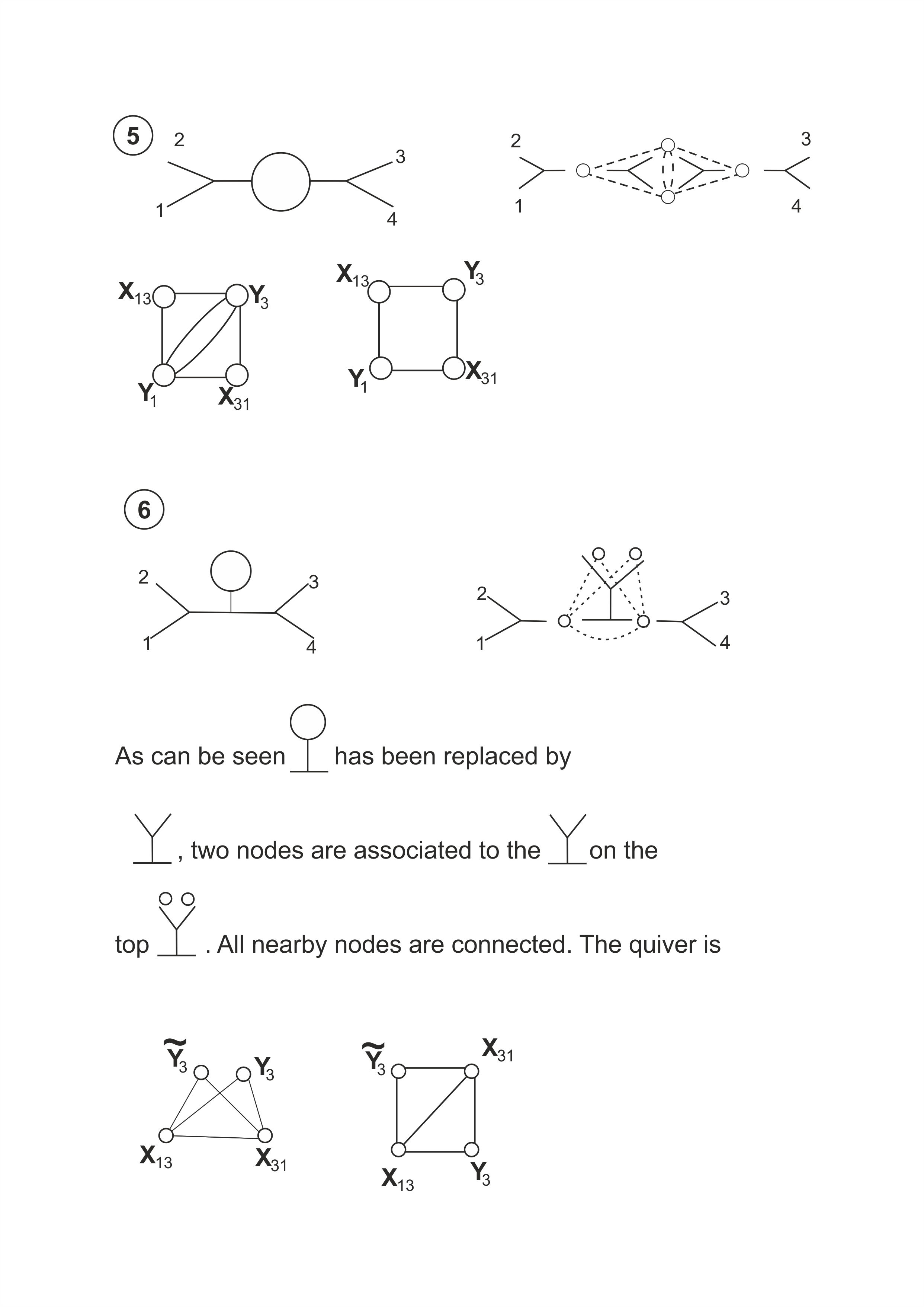}
\caption{Number 5 is a demonstration of rule 3. Number 6 is a demonstration of rule 4.}\label{1loopquiv3}
\end{figure}

In figure \ref{1loopquiv3} part 5 we see another process. Here we again decompose the Feynman diagram of the process into its $\phi^3$ vertices. We cancel out the circular part and we get the square quiver.

In part 6 we have a interesting example of a tadpole with internal lines on both sides. We again decompose the Feynman diagram into its vertices and associate two nodes with braches of the tadpole after cutting the loop open. Then, as mentioned before, we connect nodes on both sides of the tadpole to both nodes associated with the "Y" of the tadpole. This is exactly what we got for this process by connecting unfrozen nodes in \ref{All unique 1-loop processes}. Thus we have shown that using our technique we can reproduce the quivers upto 1-loops. In the next section we will construct the quivers using our technique for 2-loops and show that it matches with the quivers obtained from the other technique upto winding number 0.

\subsection{Winding number}

\begin{figure}[H]
\centering
\includegraphics[trim={0 1cm 0 1.5cm 0},clip,width=\linewidth]{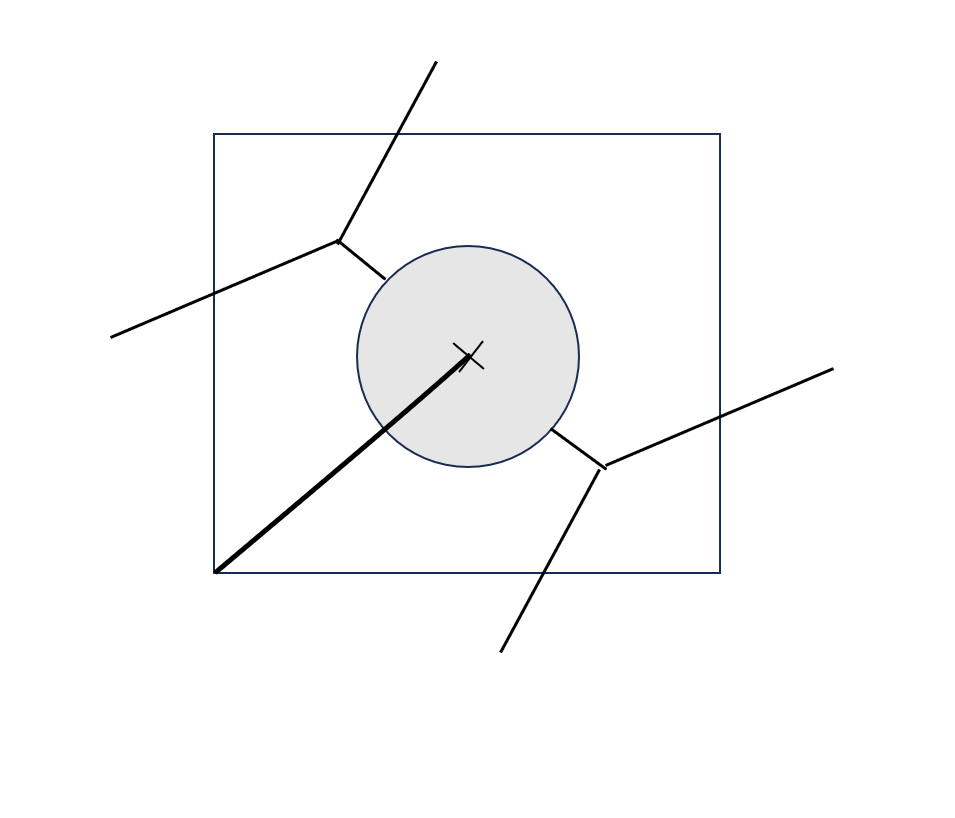}
\caption{Diagonal at winding number 0.}\label{winding1}
\end{figure}

\begin{figure}[H]
\centering
\includegraphics[trim={0 1cm 0 1.5cm 0},clip,width=\linewidth]{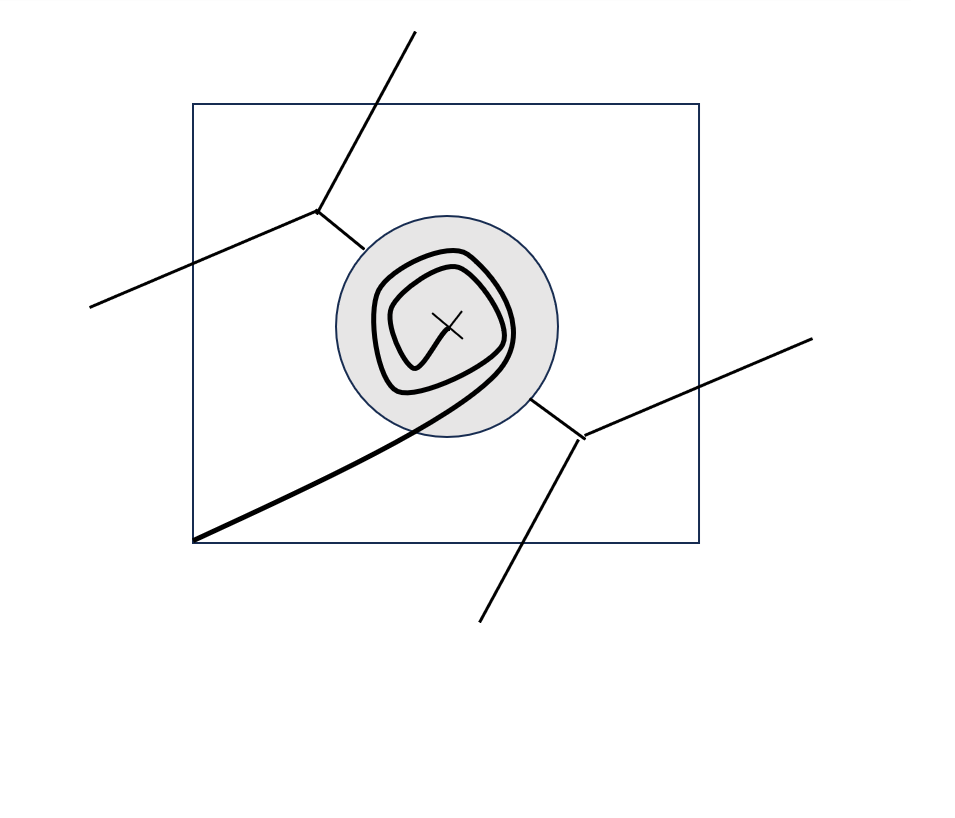}
\caption{Diagonal at winding number 2.}\label{winding2}
\end{figure}

Consider a singularity corresponding to a loop. One can draw a diagonal as shown in Fig. \ref{winding1} or one can wind around the singularity once, twice, as shown in Fig. \ref{winding2}, ... an infinite number of times and these would all be considered distinct diagonals for the same diagram. One way to see this is that topologically all such triangulations would be different. We only consider diagrams upto winding number zero in this article.

\section{Examples of quivers for two and three loop processes.}

\subsection{2-loop diagrams}
Now we will construct the quiver for all unique 2-loop processes for 4 particles for the $\phi^3$ theory. The quivers from triangulations for winding number zero can be easily obtained from known methods, just as in the case of the 1-loop processes which we will not show here.

In the figures we first show the process, then decompose the Feynman diagram into $\phi^3$ vertices and tadpoles(ref. rule 4 above). Then we draw the quivers using the construction rules. As an example we will describe the first diagram, the rest are the same.

\begin{figure}[H]
\centering
\includegraphics[trim={0 1cm 0 1.5cm 0},clip,width=\linewidth]{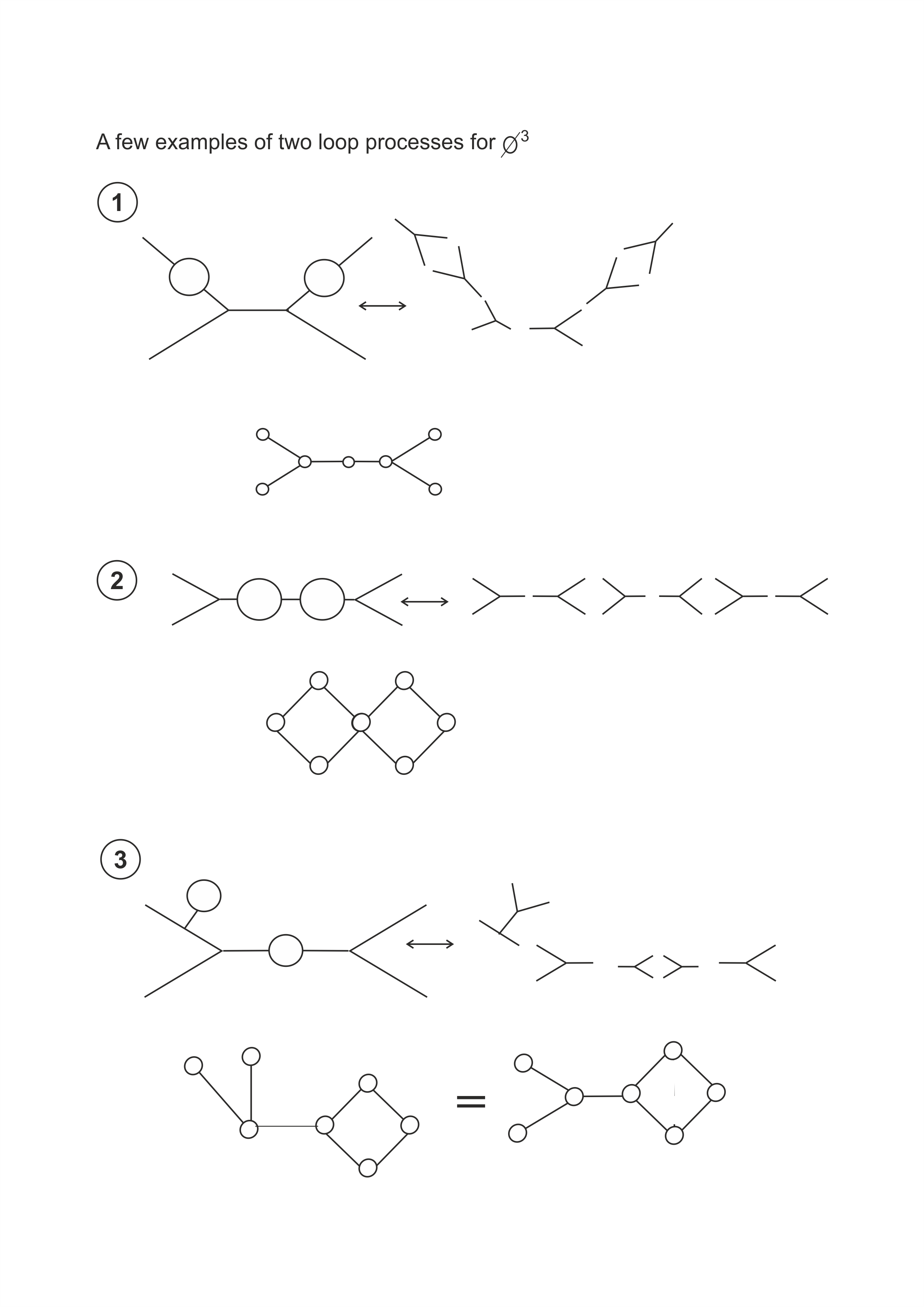}
\caption{2-loop processes for $\phi^3$.}\label{2loopquiv1}
\end{figure}

In part 1 of Fig. \ref{2loopquiv1} we start with a diagram for two leg corrections. We decompse the Feynman diagram into the $\phi^3$ vertices and assign nodes to the internal lines as described before. Then we connect all the nodes across each $\phi^3$ "vertex line". This gives us the quiver. This matches exactly with the quiver one gets from the previously known methods. We follow the same steps for the rest of the diagrams and if we draw the quiver from the known method upto winding number zero, then we will see that they all match exactly.

\begin{figure}[H]
\centering
\includegraphics[trim={0 1cm 0 1.5cm 0},clip,width=\linewidth]{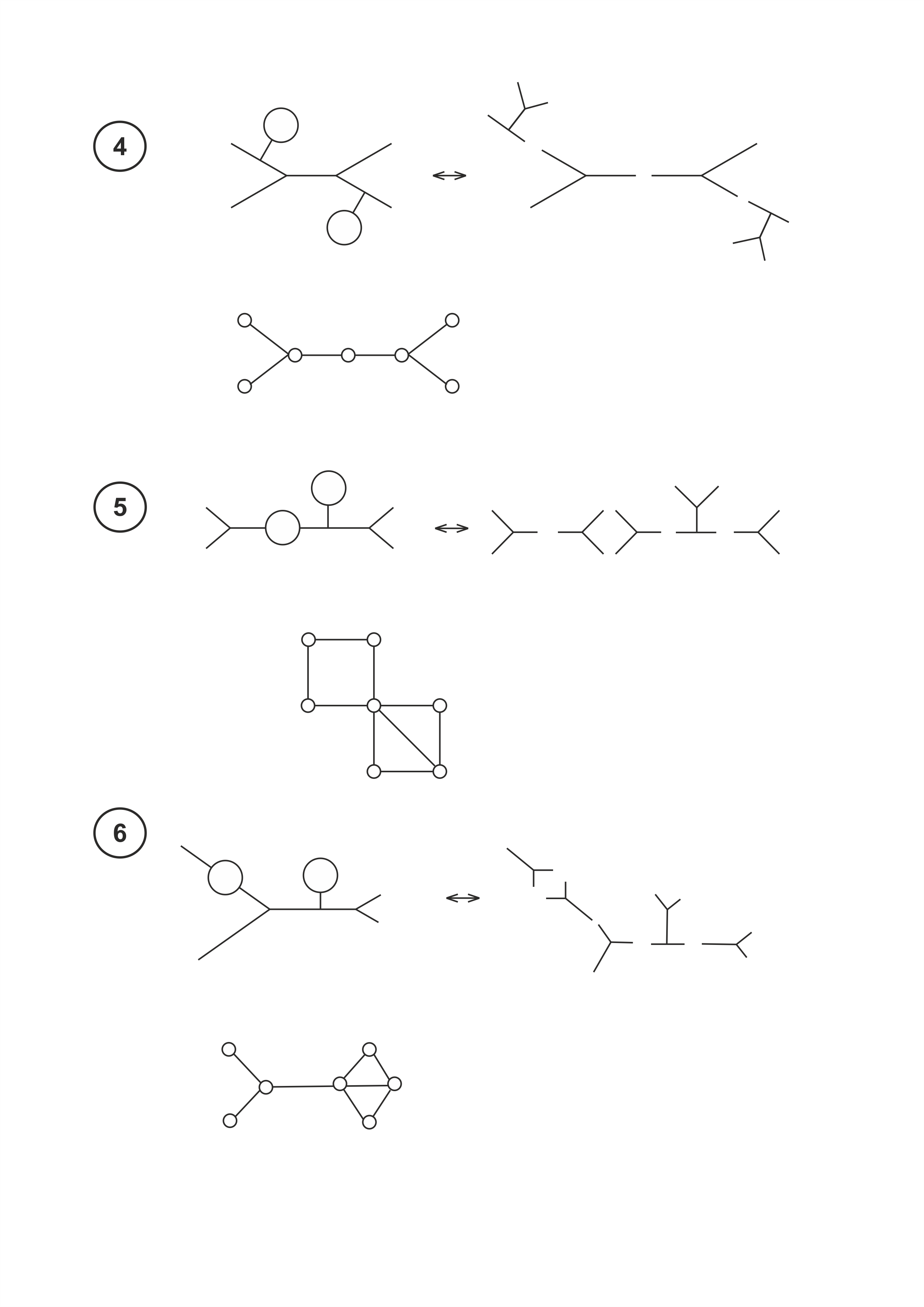}
\caption{2-loop processes for $\phi^3$.}\label{2loopquiv2}
\end{figure}

\begin{figure}[H]
\centering
\includegraphics[trim={0 1cm 0 1.5cm 0},clip,width=\linewidth]{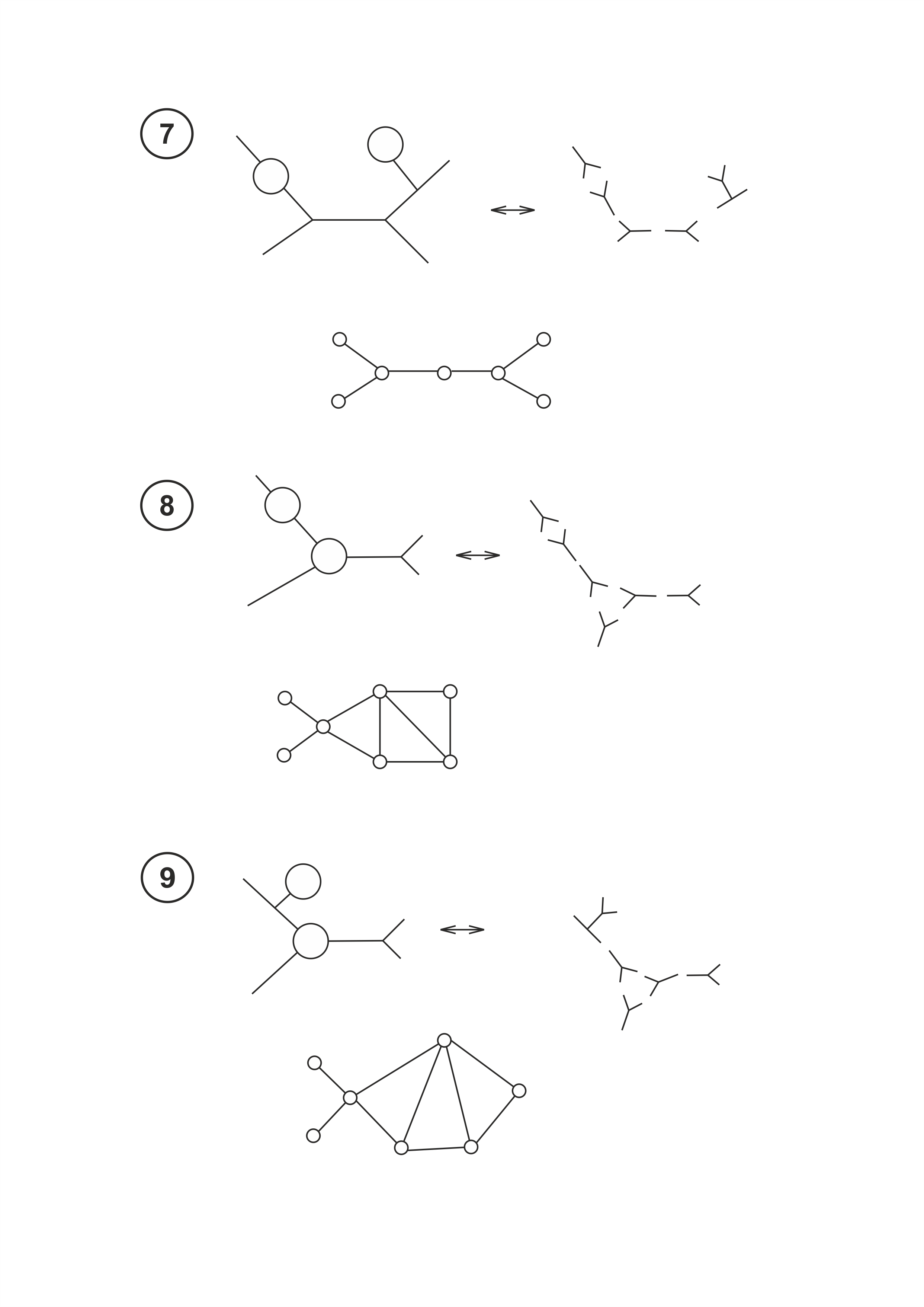}
\caption{2-loop processes for $\phi^3$.}\label{2loopquiv3}
\end{figure}

\begin{figure}[H]
\centering
\includegraphics[trim={0 1cm 0 1.5cm 0},clip,width=\linewidth]{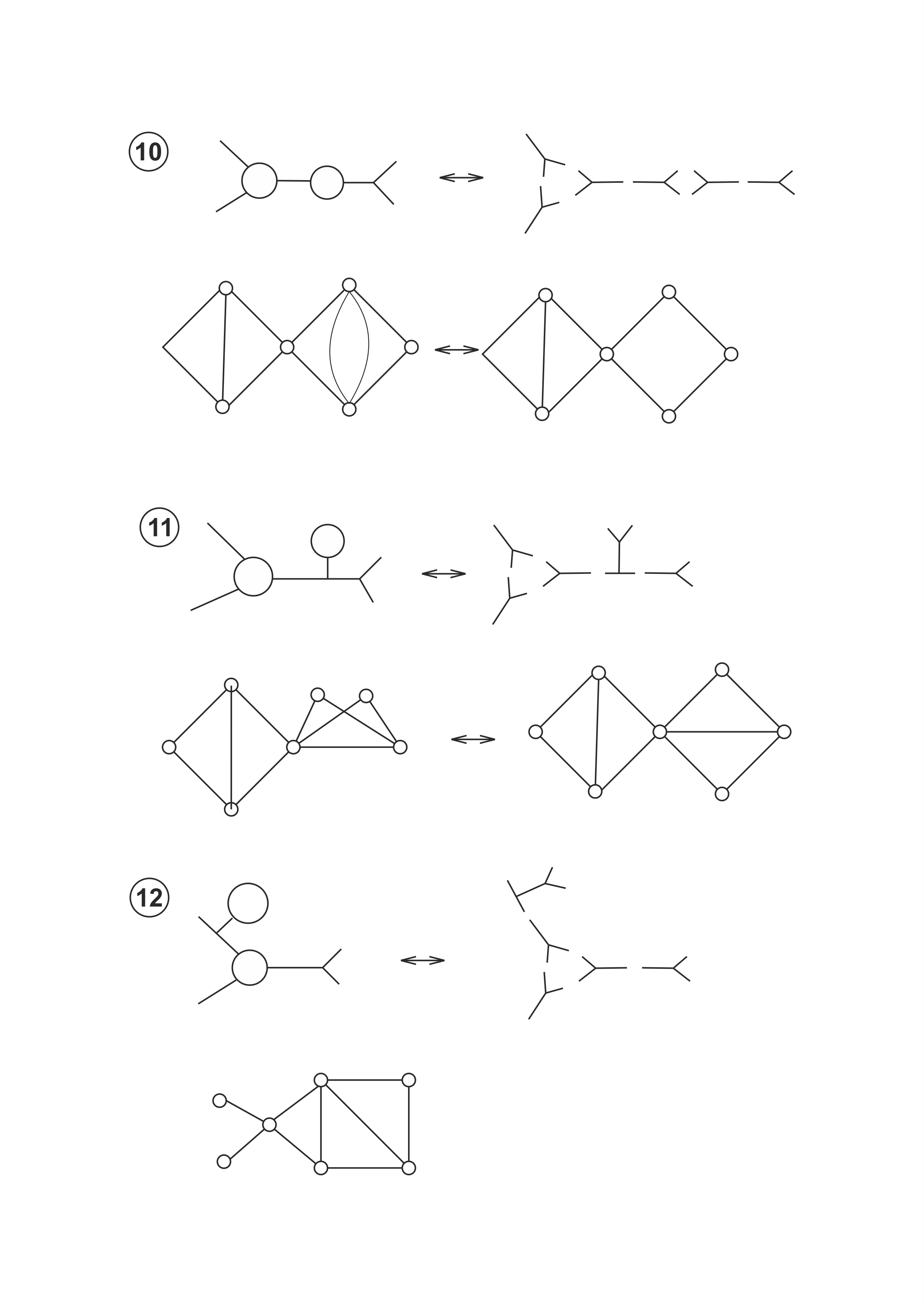}
\caption{2-loop processes for $\phi^3$.}\label{2loopquiv4}
\end{figure}

\begin{figure}[H]
\centering
\includegraphics[trim={0 1cm 0 1.5cm 0},clip,width=\linewidth]{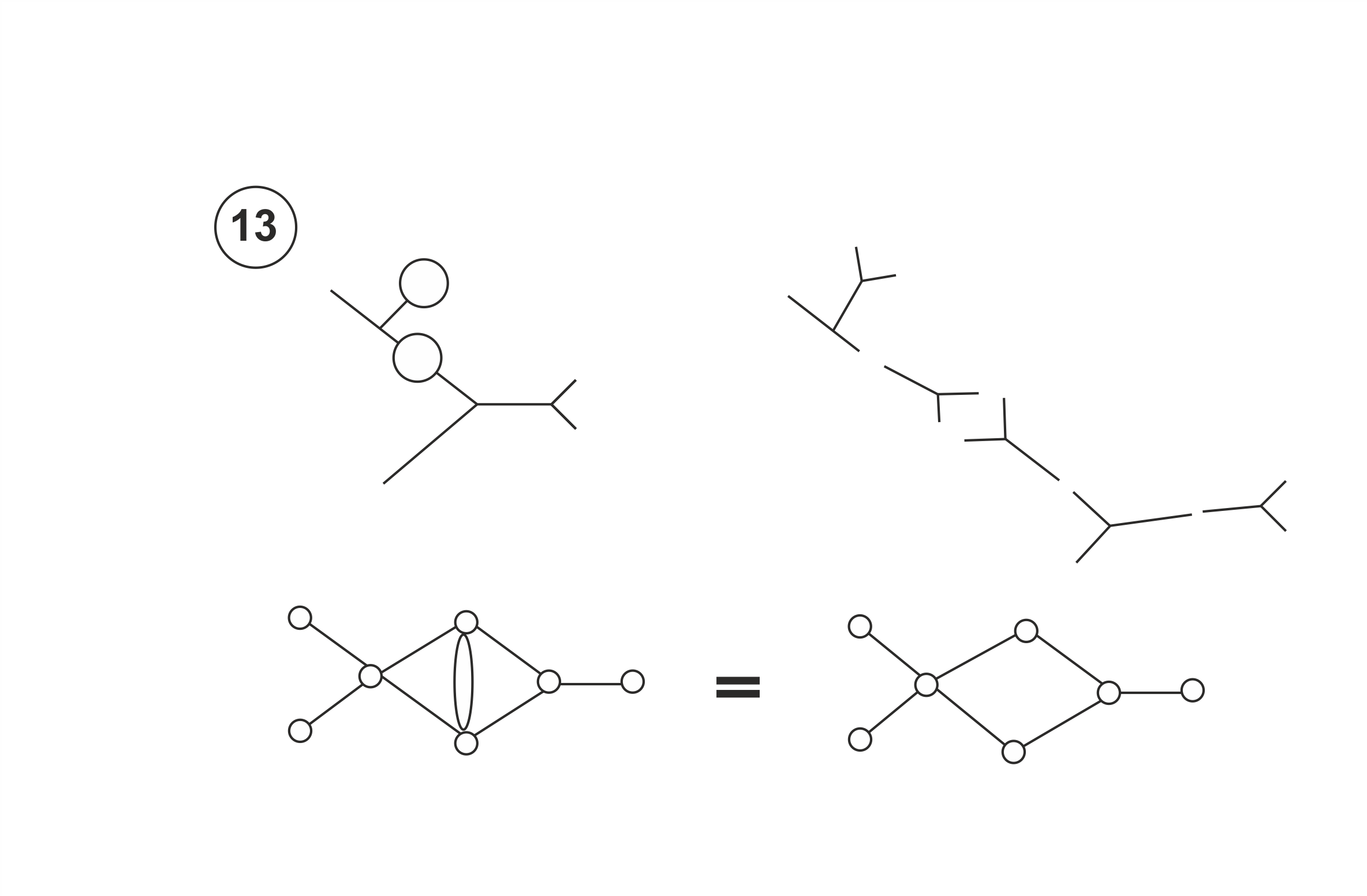}
\caption{2-loop processes for $\phi^3$.}\label{2loopquiv5}
\end{figure}

\subsection{Some non-trivial cases for 3-loop diagrams}
Here we show two processes at 3-loops for 4 particle scattering for the $\phi^3$ vertex.
\\
For both, we start with the process and then show the dual triangulation upto  winding number zero with the frozen and unfrozen nodes indicated.
\\
Next we construct the quiver using the usual method of connecting the nodes.
Then we construct the quiver using our method and show that it matches with the one before.\\

We start with the scattering process. We triangulate the dual polygon. Then mark the unfrozen nodes and connect them so that no diagonal is crossed. This way only three nodes will always be connected. This connected picture is reproduced next and is the quiver upto winding number zero for this process. Next we decompose the Feynman diagram into $\phi^3$ vertices and again assign nodes to the internal lines. Then we connect across all vertex lines. This gives us a quiver drawn on the bottom right. This matches exactly with the quiver we drew from the previous method.

We follow exactly the same steps for the next case and we again see that the quiver from our technique exactly matches the quiver from the previously known method.

\begin{figure}[H]
\centering
\includegraphics[trim={0 1cm 0 1.5cm 0},clip,width=\linewidth]{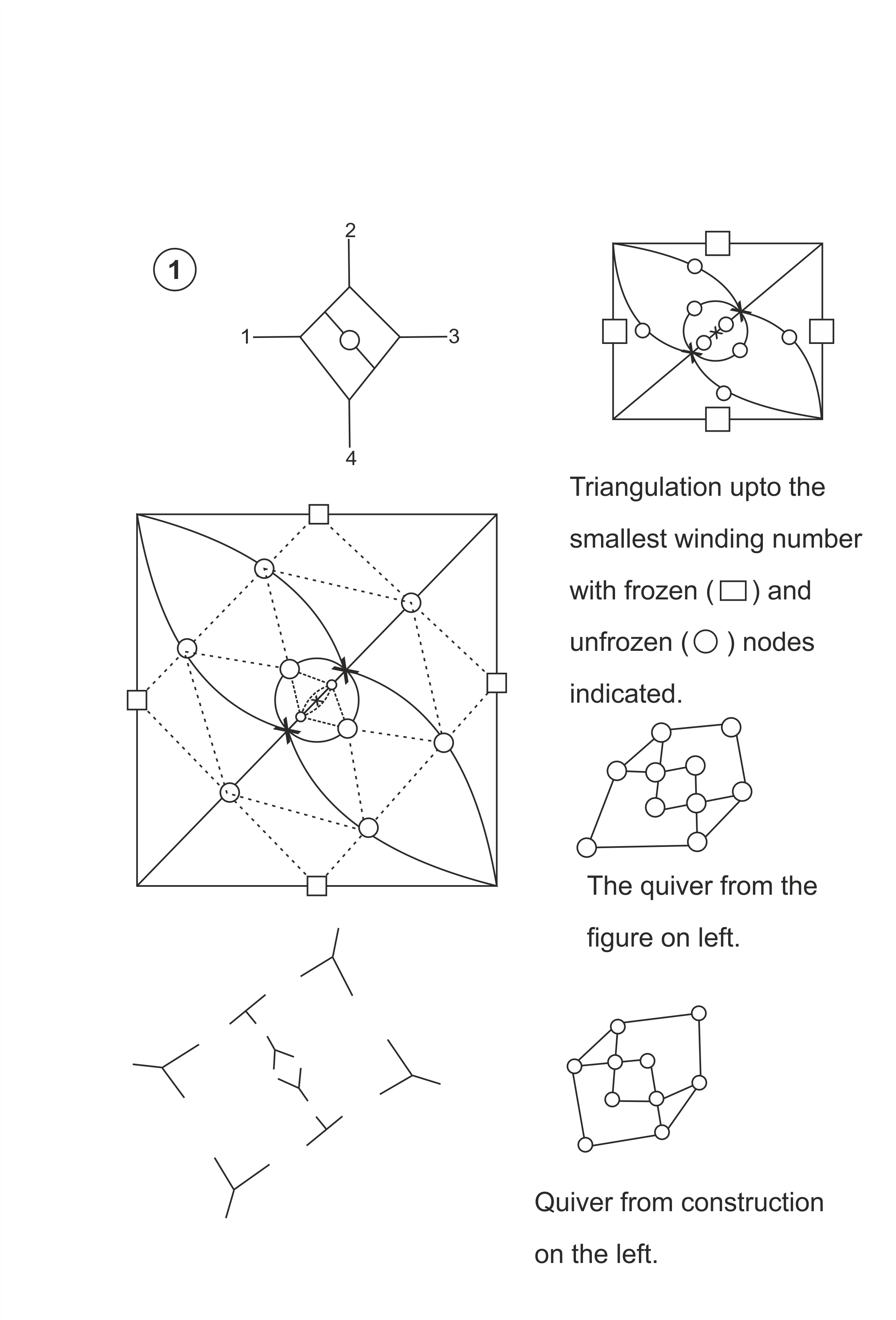}
\caption{3-loop processes for $\phi^3$ theories.}\label{3loopquiv1}
\end{figure}

\begin{figure}[H]
\centering
\includegraphics[trim={0 1cm 0 1.5cm 0},clip,width=\linewidth]{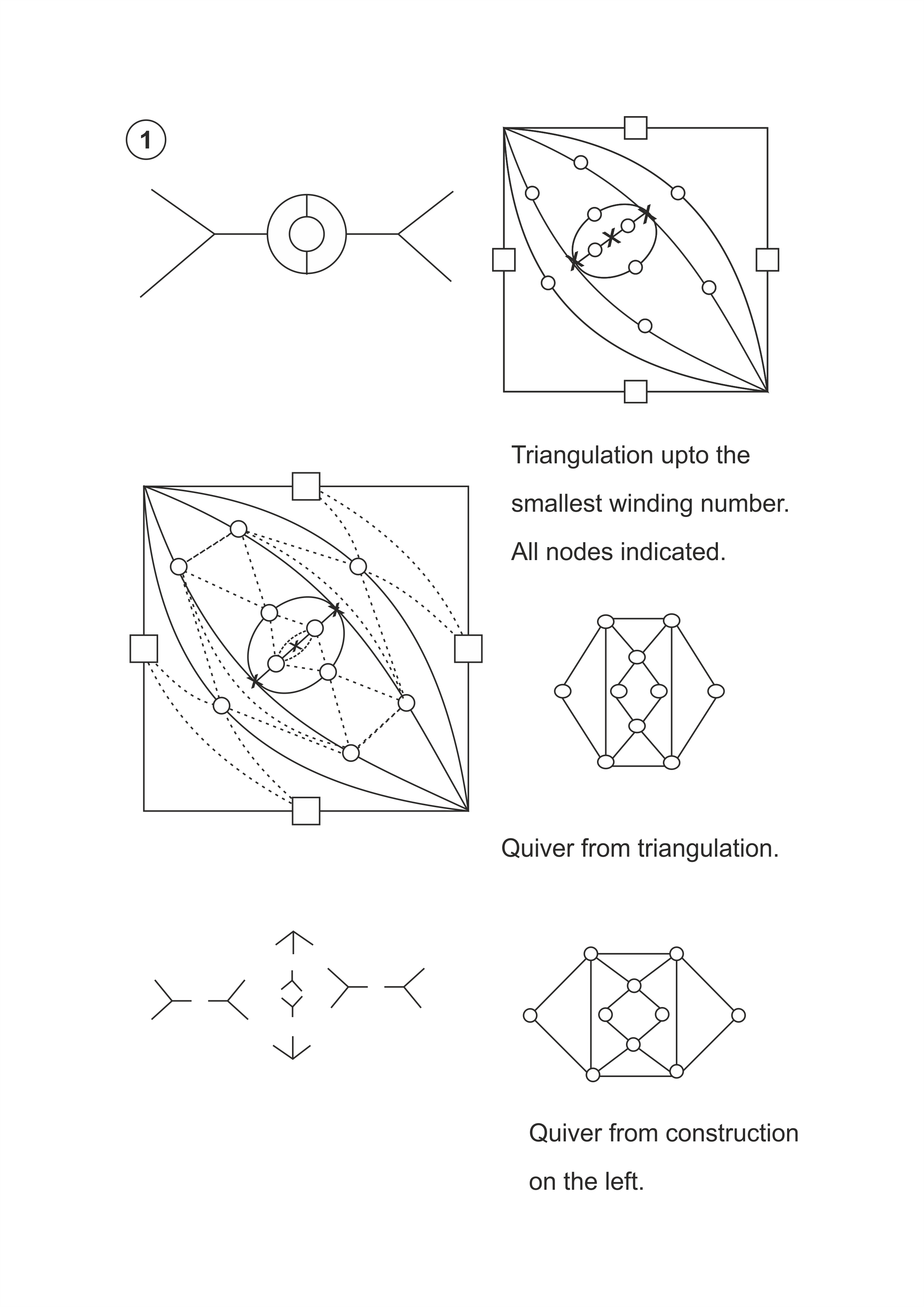}
\caption{3-loop processes for $\phi^3$ theories.}\label{3loopquiv2}
\end{figure}

\subsection{Generalization to all loop order and any particle number}

From the above examples it is clear that this program can give us the quivers for any loop order for any number of particles for winding number zero.

\section{Feynman-like rules for quivers of scattering processes for $\phi^n$ theories, $n\geq 4$.}

\subsection{Construction of quivers from polygons\\ and n-angulations(The long route).}

When writing down quivers for $\phi^3$ theories we fully triangulate the polygon and then mark the unfrozen nodes to get quivers \cite{Oak:2021lay}. This is done at all loop order for any number of particles. To get quivers for $\phi^n$ theories, for any n, we n-angulate the polygon such that each n-vertex is surrounded by n sides and diagonals.

\definecolor{xdxdff}{rgb}{0.49019607843137253,0.49019607843137253,1}
\definecolor{uuuuuu}{rgb}{0.26666666666666666,0.26666666666666666,0.26666666666666666}
\definecolor{zzttqq}{rgb}{0.6,0.2,0}
\definecolor{ududff}{rgb}{0.30196078431372547,0.30196078431372547,1}
\begin{figure}[H]
\centering
\begin{tikzpicture}[line cap=round,line join=round,>=triangle 45,x=1cm,y=1cm]
%\clip(-9.297478272524586,-4.404813668758773) rectangle (7.481972113123431,6.030406428220401);
\fill[line width=2pt,color=zzttqq,fill=zzttqq,fill opacity=0.10000000149011612] (-4.327564872585027,-0.8891193022420534) -- (-0.636085787742466,-0.8731388733033411) -- (-0.6520662166811781,2.8183402115392195) -- (-4.343545301523739,2.802359782600508) -- cycle;
\draw [line width=2pt] (-4.327564872585027,-0.8891193022420534)-- (-0.636085787742466,-0.8731388733033411);
\draw [line width=2pt] (-0.636085787742466,-0.8731388733033411)-- (-0.6520662166811781,2.8183402115392195);
\draw [line width=2pt] (-0.6520662166811781,2.8183402115392195)-- (-4.343545301523739,2.802359782600508);
\draw [line width=2pt] (-4.343545301523739,2.802359782600508)-- (-4.327564872585027,-0.8891193022420534);
\draw [line width=2pt] (-4.343545301523739,2.802359782600508)-- (-2.5856981182653773,1.0924538861582795);
\draw [line width=2pt] (-2.5856981182653773,1.0924538861582795)-- (-0.636085787742466,-0.8731388733033411);
\draw [line width=2pt] (-2.6336394050815146,0.804806165261457)-- (-2.53775683144924,1.380101607055102);
\draw [line width=2pt] (-2.8893262681009126,1.1883364597905537)-- (-2.3459916841846913,1.0445125993421425);
\draw [line width=2pt] (-5.989529482211116,0.9646104546485806)-- (-3.5125629967106957,0.45323672860978503);
\draw [line width=2pt] (-3.5125629967106957,0.45323672860978503)-- (-3.192954417936448,-2.1196123305229055);
\draw [rotate around={39.718759371554775:(-2.6416296195508706,1.2362777466066905)},line width=2pt] (-2.6416296195508706,1.2362777466066905) ellipse (1.1309025886685597cm and 0.9072122336303322cm);
\draw [line width=2pt] (-1.9080906349472169,2.0859489343532154)-- (-2.154226536920143,4.048833239820067);
\draw [line width=2pt] (-1.9080906349472169,2.0859489343532154)-- (0.8181332456803612,1.7955927594616234);
\draw (-8.336492215962877,0.16558900771296342) node[anchor=north west] {The two diagonals.};
\draw [->,line width=2pt] (-5.190508035275497,0.24549115240652433) -- (-3.7526446984122175,2.2275746504829357);
\draw [->,line width=2pt] (-5.190508035275497,0.24549115240652433) -- (-1.551097259029766,0.04937269201090233);
\draw (-6.6287466397596155,-1.784023322809945) node[anchor=north west] {The two sides.};
\draw [->,line width=2pt] (-5.46217532723361,-1.6721603202389583) -- (-4.33040207103646,-0.23372645996105912);
\draw [->,line width=2pt] (-4.838938598623827,-1.7680428938712325) -- (-3.9920513318353126,-0.8876668626717081);
\draw (-1.499028950432936,-1.5602973176679718) node[anchor=north west] {The $\phi^4$ vertex.};
\draw [->,line width=2pt] (-1.2113812295361128,-1.6721603202389583) -- (-3.512562996710696,0.4532367286097849);
\end{tikzpicture}
\caption{The $\phi^4$ vertex surrounded by two diagonals and two sides.All four sides are dual to the four external lines.}
\end{figure}

For example, let us look at the $\phi^4$ case at 1-loop for 4 particles.\\

Here, one can see that both $\phi^4$ vertices are surrounded by 2 sides and 2 diagonals = 4-gon.

We look at another example. We look at the more complicated case for 5 particles where we have a $\phi^4$ vertex and a $\phi^5$ vertex, at 1 loop.

\begin{figure}[H]
\centering
\definecolor{uuuuuu}{rgb}{0.26666666666666666,0.26666666666666666,0.26666666666666666}
\definecolor{zzttqq}{rgb}{0.6,0.2,0}
\definecolor{xdxdff}{rgb}{0.49019607843137253,0.49019607843137253,1}
\definecolor{ududff}{rgb}{0.30196078431372547,0.30196078431372547,1}
\begin{tikzpicture}[line cap=round,line join=round,>=triangle 45,x=1cm,y=1cm]
%\clip(-8.439347739941596,-6.61372920465999) rectangle (9.736853855188333,4.690137120692235);
\fill[line width=2pt,color=zzttqq,fill=zzttqq,fill opacity=0.10000000149011612] (-0.774588691202375,-4.0078931918923955) -- (3.9715987035952023,-3.959951905076259) -- (5.39265639384975,0.5687552166622831) -- (1.5247309516039047,3.319708856174276) -- (-2.2868361279090452,0.4911845851293697) -- cycle;
\draw [line width=2pt] (-6.996243831430383,-0.7796358505940455)-- (-5.893594234659228,-1.9142463052426235);
\draw [line width=2pt] (-5.893594234659228,-1.9142463052426235)-- (-6.8684003999206835,-2.777189467933092);
\draw [rotate around={0:(-5.462122653313992,-1.8982658763039115)},line width=2pt] (-5.462122653313992,-1.8982658763039115) ellipse (0.43867801219087027cm and 0.40191806526061985cm);
\draw [line width=2pt] (-5.023451589995231,-1.896003655296266)-- (-4.647120777439662,-0.7956162795327575);
\draw [line width=2pt] (-5.023451589995231,-1.896003655296266)-- (-4.151727480339578,-1.8183637316103496);
\draw [line width=2pt] (-5.023451589995231,-1.896003655296266)-- (-4.487316488052538,-2.649346036423391);
\draw [line width=2pt,color=zzttqq] (-0.774588691202375,-4.0078931918923955)-- (3.9715987035952023,-3.959951905076259);
\draw [line width=2pt,color=zzttqq] (3.9715987035952023,-3.959951905076259)-- (5.39265639384975,0.5687552166622831);
\draw [line width=2pt,color=zzttqq] (5.39265639384975,0.5687552166622831)-- (1.5247309516039047,3.319708856174276);
\draw [line width=2pt,color=zzttqq] (1.5247309516039047,3.319708856174276)-- (-2.2868361279090452,0.4911845851293697);
\draw [line width=2pt,color=zzttqq] (-2.2868361279090452,0.4911845851293697)-- (-0.774588691202375,-4.0078931918923955);
\draw [line width=2pt] (-2.2868361279090452,0.4911845851293697)-- (1.2373157759418605,-0.7627233578467398);
\draw [line width=2pt] (1.2373157759418605,-0.7627233578467398)-- (3.9715987035952023,-3.959951905076259);
\draw [line width=2pt] (1.2200051077560223,-1.1608687261210142)-- (1.2200051077560223,-0.2953353168291134);
\draw [line width=2pt] (0.7353063985525576,-0.7454126896609018)-- (1.600839807844459,-0.7627233578467398);
\draw [rotate around={62.44718842328246:(1.0122770895259663,-0.6934806851033879)},line width=2pt] (1.0122770895259663,-0.6934806851033879) ellipse (1.1763296001343098cm and 1.087235330697132cm);
\draw [line width=2pt] (-2.7441379068008858,-1.8706061217403729)-- (0.4526216152830089,-1.7281227877066783);
\draw [line width=2pt] (0.4526216152830089,-1.7281227877066783)-- (1.0815197622693182,-5.12501174067792);
\draw [line width=2pt] (1.5587346665780593,0.3482170960382154)-- (-1.5323911337922238,3.5995650249844404);
\draw [line width=2pt] (1.5587346665780593,0.3482170960382154)-- (4.387857385764382,3.287972997639356);
\draw [line width=2pt] (1.5587346665780593,0.3482170960382154)-- (5.530361486029691,-1.2820434034218804);
\draw (-5.16763145281821,1.0626248198963081) node[anchor=north west] {The $\phi^4$ vertex.};
\draw [->,line width=2pt] (-3.367321961491055,0.3278487378610552) -- (-0.06098433799599157,-1.5590140943952884);
\draw (-0.6495470563144845,4.188127743302933) node[anchor=north west] {The $\phi^5$ vertex.};
\draw (-5.842747512065893,2.370507583789941) node[anchor=north west] {\parbox{5.016075019114422 cm}{The singularity corresponding  to the loop.}};
\draw (-6.864076935030337,-3.5497409357666605) node[anchor=north west] {The scattering process.};
\draw [->,line width=2pt] (0.4583357075791492,3.236040993081842) -- (1.514286466915269,0.6567514333919777);
\draw [->,line width=2pt] (-2.190196524854069,1.7819448654714485) -- (1.0988304304551564,-0.5376846714308456);
\end{tikzpicture}
\caption{Scattering amplitude for 5 particles where we have a $\phi^4$ vertex and a $\phi^5$ vertex, at 1 loop.}
\end{figure}

As can be seen, the $\phi^4$ vertex is surrounded by 2 sides of the polygon + 2 diagonals = 4. Similarly the $\phi^5$ vertex is surrounded by 3 sides + 2 diagonals = 5-gon.

Thus, an n-gon for a $\phi^n$ vertex.

Now we describe how to obtain quivers using the dual polygon and its n-angulations and look at an example.

\begin{enumerate}

\item
Assign an unfrozen node where the diagonals and internal propagators intersect.

\item
Assign frozen nodes where the externals lines and the sides of the polygon intersect.

\item\label{3}
The lines connecting the nodes are called edges. Connect all nodes such that the edges of the quiver never cross external lines.

\item\label{4}
Connect all nodes such that the edges of the quiver never cross internal propagators.

\item\label{5}
Connect all nodes such that the edges of the quiver never cross diagonals.

\begin{figure}[H]
\centering
\includegraphics[scale=.55]{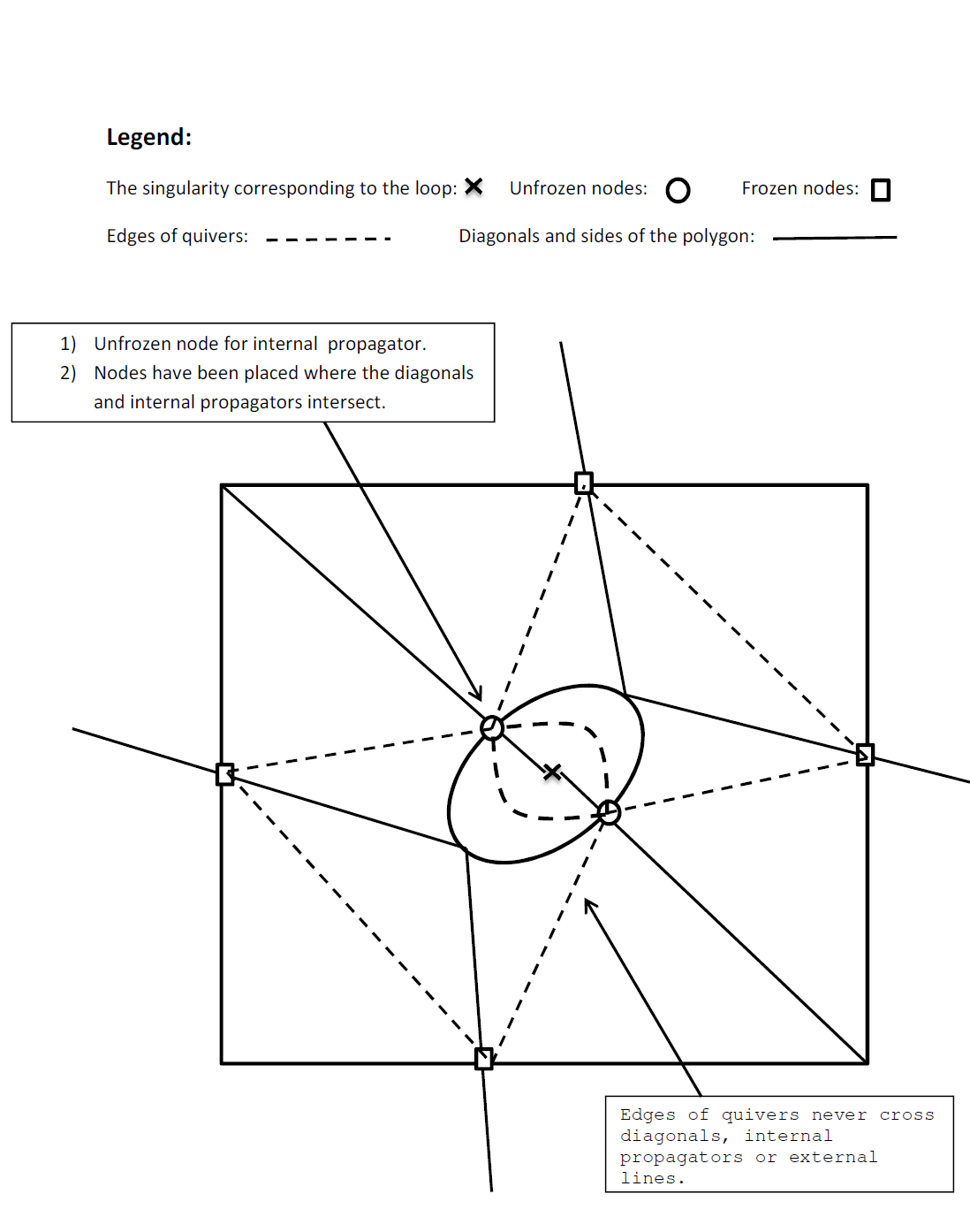}
\caption{Illustration of quiver construction for 4 point 1-loop scattering process for $\phi^4$ theory.}
\end{figure}

\item
Note down all the unfrozen nodes and the connections between them. Ignore all edges involving unfrozen nodes.

\end{enumerate}

\begin{figure}[H]
\centering
\includegraphics[trim={0 16cm 0 4},clip=true,width=\linewidth]{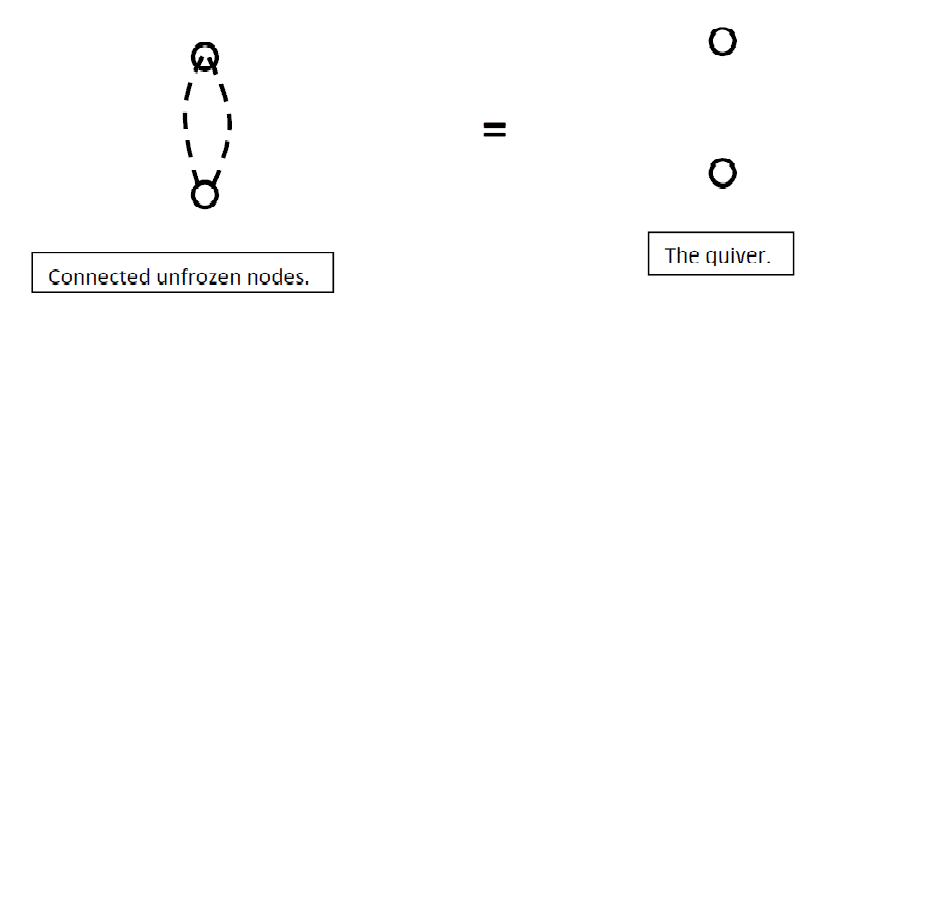}
\caption{All unfrozen nodes and the edges between them and the final quiver.}
\end{figure}

\subsection{The need for the new approach.}

At the tree level, it seems, the scattering amplitude for any number of particles for any interaction $\phi^n$ can be computed using a computer program. It would be ideal if such a program can be made for the entire process. The traditional method is to use the Feynman diagrammatic expansion. The automation by the computer only addresses the second half of the problem of computing the scattering forms once the quivers are known. To do their computation, the old approach of quiver construction, is to draw the dual polygon, n-angulate, draw all nodes, frozen and unfrozen, and connect them. This is a long and an arduous procedure which also does not seem to yield to an algorithmic(in the sense of being automated using a computer program) approach. The new rules based approach might possibly be implemented in a computer program.

  Another point of interest, if both the approaches of computing the quivers are implemented using a computer program, the latter approach has a much lesser number of steps. This means the latter approach is more efficient computationally and would yield results much faster.

\subsection{The Feynman-like rules for constructing quivers.}

Now we list out a few basic structures using which all quivers for any theory, $\phi^n$ for $n\geq4$, at all loop order, can be constructed.

$n=3$ is a special case about which we will make a few comments later.

We will now draw the basic structures. Then we will list out the rules for drawing quivers from Feynman diagrams without dual polygons and n-angulations. A step by step construction of quivers will be demonstrated after that.

\begin{figure}[H]
\centering
\includegraphics[trim={0 0cm 0 0cm}, clip,width=\linewidth]{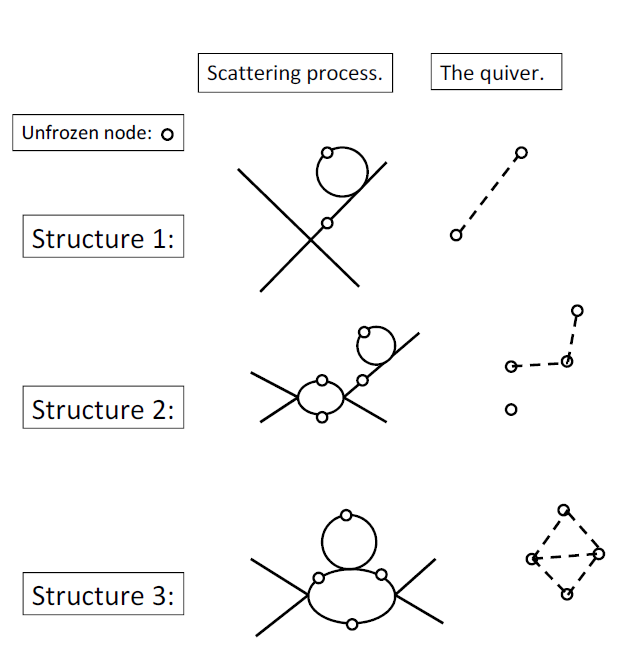}
\caption{Basic structures using which all quivers for any theory, $\phi^n$ for $n\geq4$, at all loop order, can be constructed.}
\end{figure}

\subsection{The rules.}

\begin{figure}[H]
\centering
\includegraphics[trim={0 7cm 0 0}, clip,width=\linewidth]{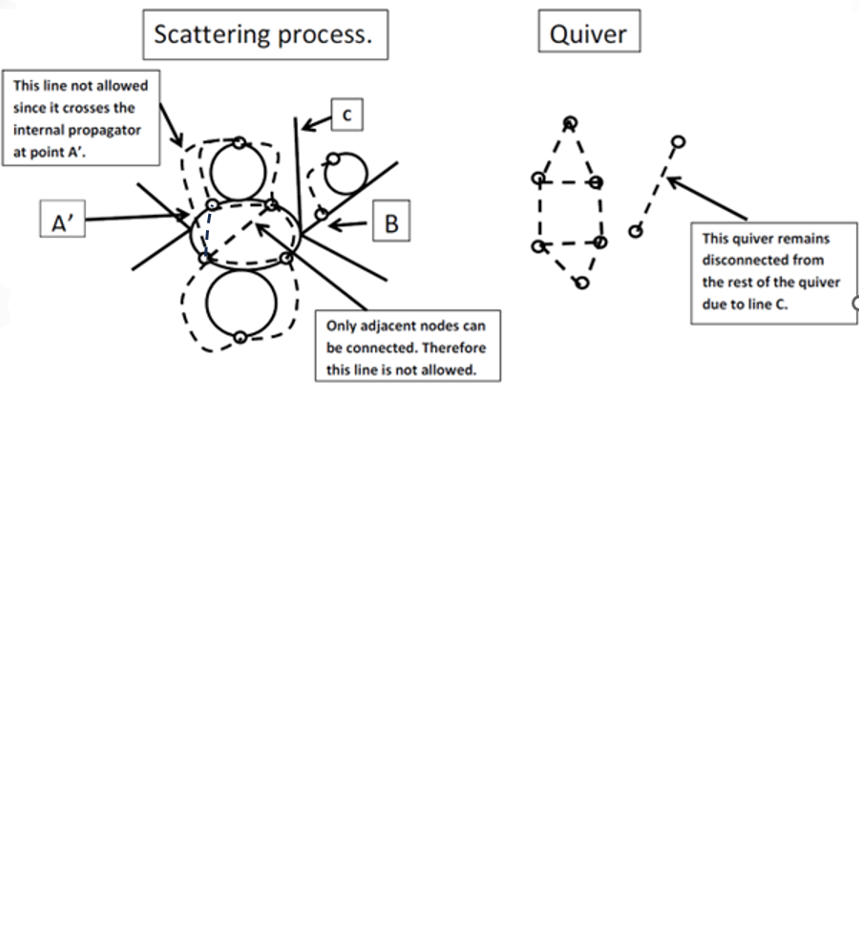}
\caption{A 5-point scattering amplitude at four loop.}\label{equivRule4diag1mod}
\end{figure}

\begin{enumerate}

\item
Assign an unfrozen node to all internal propagators.

\item

We look at the two figures below. Figure \ref{equivRule3fig1} is a 5-point scattering amplitude at two loops. There is no external line between unfrozen nodes A and B. Therefore, they can be connected. Figure \ref{equivRule3fig2} is a 5-point scattering amplitude at two loops. There is an external line between unfrozen nodes A and B. Therefore, they cannot be connected. This implements rule \ref{3}.

\begin{figure}[H]
\centering
\includegraphics[trim={0 10cm 0 0}, clip,width=\linewidth]{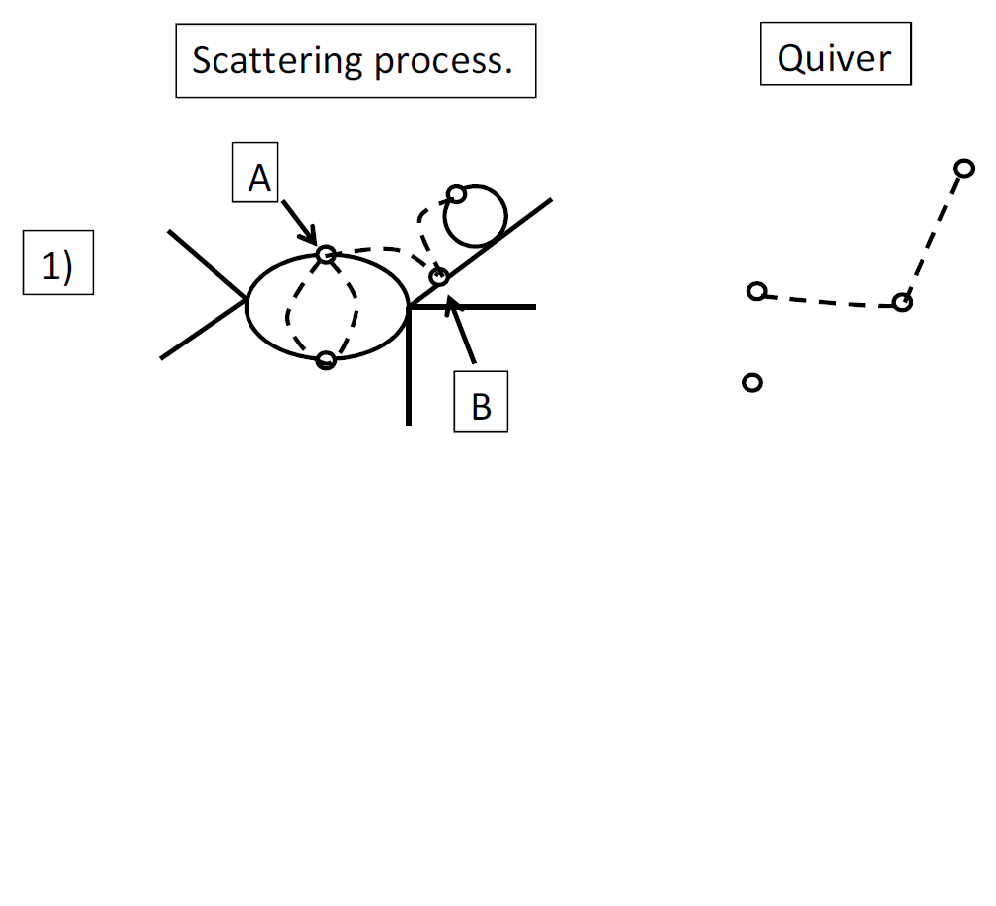}
\caption{A 5-point scattering amplitude at two loop. No external line between unfrozen nodes A and B. Therefore, they can be connected.}\label{equivRule3fig1}
\end{figure}

\begin{figure}[H]
\centering
\includegraphics[trim={0 10cm 0 0}, clip,width=\linewidth]{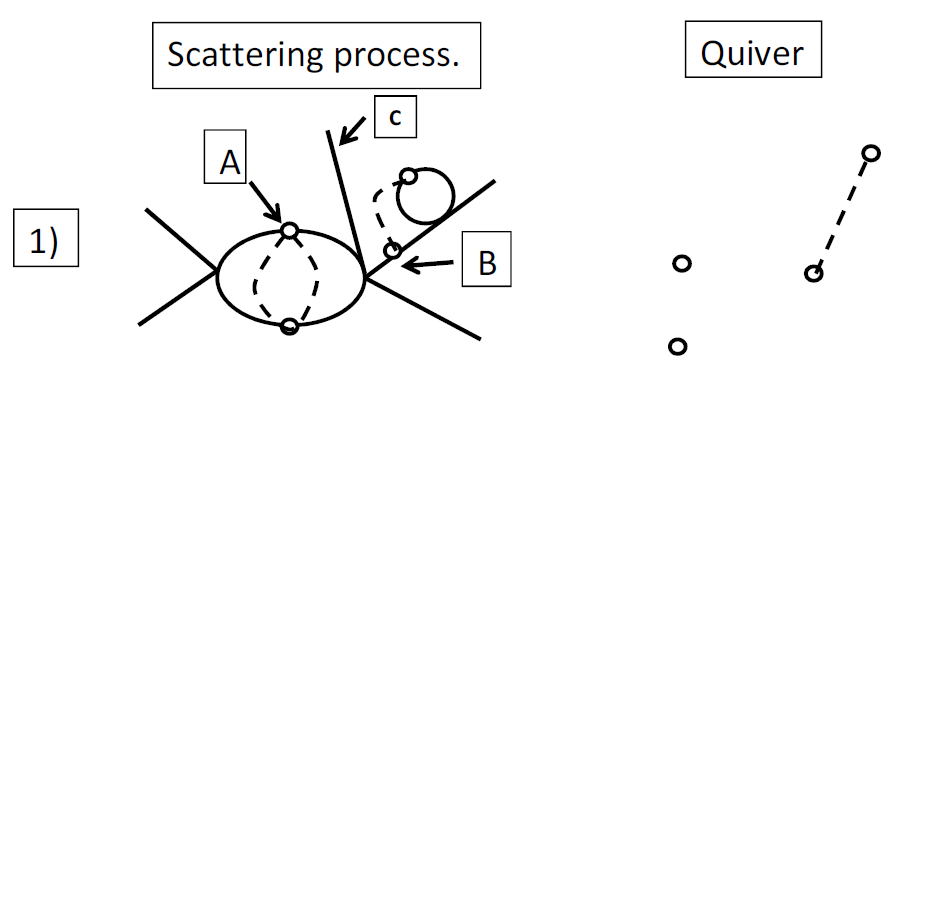}
\caption{A 5-point scattering amplitude at two loop. There is an external line marked 'C' between the unfrozen nodes A and B. This prohibits the connection between the unfrozen nodes 'A' and 'B'.}\label{equivRule3fig2}
\end{figure}

\item
As shown in figure \ref{equivRule4diag1mod}, the edge of the quiver crossing the internal propagator at the point A' is prohibited.
This implements rule \ref{4}.

\item
Again, as shown in figure \ref{equivRule4diag1mod}, the edge of the quiver connecting two non-adjacent nodes is prohibited.
This implements rule \ref{5}.

\end{enumerate}

Using the above heuristic rules we believe we can write down the quivers for any scattering process, for any number of particles, for any theory $\phi^n$, $n\geq4$, at all loop order.

\subsection{The $\phi^3$ case.}

In section \ref{1loopquivers}, in the $\phi^3$ case, we saw the need for a special rule, \hyperlink{$\phi^3$ rule 4.}{Rule 4} for tadpoles. We find that there is no need for such a rule in any other case. This rule for tadpoles is specific only to this case.

\section{Summary and Conclusions}

In this paper, we begin by explaining the need for the computation of the scattering amplitude of any process, namely, that they are the only observables of quantum gravity in asymptotically flat space-time. Then we explain the need of a new approach to the problem of computing scattering amplitudes, that, a first principles derivation of the fundamental analyticity properties encoding unitarity and causality of the theory of scattering amplitudes, failed. In the next section we introduce the "Amplituhedron" approach started by Nima Arkani-Hamed et al for coloured scalar theories. Then we expand on the problem of computing scattering amplitudes for these coloured scalar theories. In section 3 we review the known results for quivers for tree level and 1-loop cases. In the next section we motivate and prescribe a set of Feynman-like rules to construct quivers from Feynman diagrams directly. In section 5 we demonstrate the construction of quivers for 2-loop and 3-loop processes. After that we juxtapose the two approaches of computing the quivers. We also demonstrate the construction of quivers using our approach, step by step. Then we see the construction of quivers for a few non-trivial cases of scattering processes. In the next section we list the rules of construction for quivers for $\phi^n$ theories, $n\geq 4$. It can be easily seen from here that this program can be generalized to all loop orders for all theories, including mixed, for any number of particles.

\section*{Future directions}

\cite{Barmeier:2021iyq} have computed the scattering amplitude for tree level processes for any number of particles. They start with the quiver of the tree level process, write down its Auslander-Reiten walk and then compute the amplitude. Here the quiver is 1-dimensional and the Auslander-Reiten walk is 2-dimensional. For computing amplitudes in any theory $\phi^n$ for 1-loop and above, the quiver is 2-dimensional, therefore the AR walk will be 3-dimensional. We hope to address that in our next paper and compute the scattering amplitude thereof. Also, note that the AR walk is 3-dimensional and no more for any loop order for any theory.

\section*{Acknowledgments}

We are deeply indebted to Koushik Ray for the initial idea of this program. We thank Mr. Bharat Bhushan for teaching us how to generate the figures on the computer.

\end{document}